\documentclass[11pt]{article}
\usepackage{notations}
\usepackage{url}
\usepackage[utf8]{inputenc}
\usepackage[letterpaper,margin=1.0in]{geometry}
\usepackage{authblk}
\usepackage[numbers]{natbib}
\usepackage{microtype}
\begin{document}
\title{Robust Node Localization for Rough and Extreme Deployment Environments}
\date{\vspace{-4ex}}

\author[1]{Abiy Tasissa}
\author[2]{Waltenegus Dargie}
\affil[1]{Department of Mathematics, Tufts University, Medford, MA 02155, USA.
}
\affil[2]{Faculty of Computer Science, Technische Universität Dresden, Dresden, Germany. 
}

\maketitle

\begin{abstract}

Many applications have been identified which require the deployment of large-scale low-power wireless sensor networks. Some of the deployment environments, however, impose harsh operation conditions due to intense cross-technology interference, extreme weather conditions (heavy rainfall, excessive heat, etc.), or rough motion, thereby affecting the quality and predictability of the wireless links the nodes establish. In localization tasks, these conditions often lead to significant errors in estimating the position of target nodes. Motivated by the practical deployments of sensors on the surface of different water bodies, we address the problem of identifying susceptible nodes and robustly estimating their positions. We formulate these tasks as a compressive sensing problem and propose algorithms for both node identification and robust estimation. Additionally, we design an optimal anchor configuration to maximize the robustness of the position estimation task. Our numerical results and comparisons with competitive methods demonstrate that the proposed algorithms achieve both objectives with a modest number of anchors. Since our method relies only on target-to-anchor distances, it is broadly applicable and yields resilient, robust localization.

\end{abstract}

\section{Introduction}
\label{sec:intro}

Low-power wireless sensor networks enable cost-effective and scalable monitoring of different physical environments \citep{dargie2023monitoring, sanjeevi2020precision}. The networks can achieve high spatio-temporal sensing if the precise location of the nodes can be determined. For static deployments, the location can be established during deployment with relative ease. If the nodes are mobile or quasi-mobile, however, their location has to be determined/updated on-the-fly. The difficulty of this task depends on many factors, including the environment dynamics. Some of the deployment environments impose harsh operation conditions (strong wind, excessive heat, excessive rain, rough motion, etc.), making localization challenging. Integrating sophisticated localization systems may not be a sustainable option for many of the networks due to cost and energy consumption. The most feasible approaches aim to exploit some models of the deployment environments to predict signal propagation and to use this knowledge for localization \citep{Wang10054103}. For line-of-sight (LOS) wireless channels, for example, the received power is a function of the inverse of the square of the distance separating a receiver from a transmitter. Often this knowledge is sufficient to determine the distance from the received power. When the environment is rough and the nodes are subject to various types of motions, the LOS model is no longer sufficient, in which case other models, such as two-ray \citep{weiss2022semi}, three-ray  \citep{hoydis2023sionna}, or more advanced models should be applied \citep{wang2018wireless, lei2009multipath}. Similarly, localization techniques such as Time Difference of Arrival (TDOA) and Time of Flight (TOF) benefit from models of the deployment environments to compensate for factors which affect signal propagation \citep{shamaei2021receiver, zhao2021learning}

In this paper we propose a resilient localization model for wireless sensor networks deployed in extreme and rough environments. In the networks of interest, we assume $n$ mobile or quasi-mobile nodes are randomly distributed around $m$ anchor nodes. The positions of the anchor nodes are known, while the positions of the mobile nodes are unknown. A significant number of the mobile nodes are relatively close to the anchor nodes and their relative distance can be established with acceptable certainty (e.g., with an accuracy exceeding 90\%), by applying appropriate path loss models. However, a subset of the mobile nodes is far from the anchor nodes and their relative distance can be established only to a certain degree of accuracy ($\leq$ 70\%). \\

\noindent \textbf{Contributions}: Our model offers three key contributions:
\begin{enumerate}
    \item The proposed model  dynamically identifies the second types of nodes which suffer from a considerable amount of uncertainty in their distance estimation.

\item It quantifies the error embedded in the distance of the individual nodes and extracts it from the distance estimation, enabling accurate localization of the mobile nodes. 

\item We derive conditions on the optimal deployment configurations that facilitate robust estimation of highly corrupted mobile nodes. 
\end{enumerate}
The remaining part of the paper is organized as follows. In Section \ref{sec:related}, we review related work. In Sections \ref{sec:motivation} and \ref{sec:background}, we provide the motivation for the paper and background information in trilateration and compressive sensing.  In Section~\ref{sec:setup} we formulate the problem statement and identify the principal tasks the paper sets out to accomplish. In Sections~\ref{sec:estimation} and \ref{sec:identification}, we present our strategy to estimate the position of target nodes and to identify severely corrupted nodes. This is followed by a theoretical discussion of the conditions under which our proposed algorithms yield accurate identification and estimation. In Section~\ref{sec:config},  we discuss our strategy for optimally configuring deploying parameters. In Section~\ref{sec:experiments} and \ref{sec:discussion}, we present numerical experiments and discuss results. Finally in Section~\ref{sec:conclusion} we provide concluding remarks and outline future work.

\section{Related Work}
\label{sec:related}

Localizing low-power, resource-constrained sensing nodes remains an active research challenge. Proposed approaches vary based on several key factors, including the scale and distribution of the deployment, the nature of the anchor nodes, and the relationship between mobile (target) and anchor nodes (useful for identifying the appropriate optimization models and introducing realistic constraints). Additionally, approaches differ in how they account for the mobility of target nodes and the error statistics associated with distance measurements between anchor and target nodes, among other considerations.  

Our work formulates the localization problem as a compressive sensing problem, and as such has connections to  sparse error correction and structured compressive sensing \citep{candes2008highly,candes2005error,eldar2010block,baraniuk2010model}. We will provide the necessary background for compressive sensing in Section~\ref{sec:compressed_sensing}. In what follows, we briefly review related work in robust localization.

A seminal approach for outlier detection is the robust PCA framework \citep{candes2011robust}, which decomposes a matrix into a low-rank component and a sparse outlier matrix. This framework can be used to identify susceptible target nodes from corrupted distance measurements. In our case, the squared distance matrix between target nodes and anchors (of size $m \times n$) is low-rank \citep{dokmanic2015euclidean}, enabling the application of robust PCA. Moreover, when the corruptions are structured, one can consider structured variants of robust PCA \citep{tang2011robust,liu2015background}. These methods are well developed, come with theoretical guarantees \citep{candes2011robust}, and efficient algorithms exist. However, they typically perform well when $m$ (the number of anchors) is relatively large, a condition that is ideal for big-data problems but limiting when the number of anchors is constrained. We empirically demonstrate the limitation of the structured robust PCA approach in the few anchors regime in Section~\ref{sec:experiments}.

In \citep{zaeemzadeh2017robust}, the authors address robust target localization by modeling each distance measurement $r_i$ from an anchor $a_i$ to a target node at $x$ as
\[
r_i = \|\bm{x} - \bm{a}_i\|^2 + \nu_i, \quad i=1,\dots,m,
\]
where the measurement error is given by
\[
\nu = (1-\beta)\,\mathcal{N}(0,\sigma^2) + \beta\,\mathcal{H}(\nu).
\]
Here, $\mathcal{N}(0,\sigma^2)$ represents Gaussian noise (i.e., outlier-free error) and $\mathcal{H}(\nu)$ denotes the outlier error. Using techniques from robust statistics \citep{Huber2011} and iteratively reweighted least squares minimization \citep{daubechies2010iteratively}, they propose two algorithms for this task. We compare our method with their approach in the numerical results section.

Another related approach is presented in \citep{li2017outlier}, which proposes a robust localization method based on non-convex robust PCA. This work falls within the framework of device-free localization, where the target area is discretized into a grid and equipped with a set of transmitting and receiving anchor nodes. For each grid point, representing a potential target location, the received signal strength (RSS) between pairs of transmitting and receiving anchors is recorded \citep{zhang2023device,youssef2007challenges}. At the online stage, the goal is to predict the target's location using the pre-collected RSS profiles. The work in \citep{feng2011received} addresses a similar localization problem using RSS measurements in indoor settings via compressive techniques. Our approach differs from these works in a crucial way. In our motivating application (e.g., water quality monitoring), the target area is vast, rendering grid discretization and comprehensive RSS profiling expensive.

In \citep{moore2004robust}, the authors formulate the localization problem as a graph realization problem and develop a robust, distributed algorithm to localize a node using only distance measurements to nearby sensor nodes. We note that while this method operates without anchors, it requires distance measurements between nearby target nodes. In contrast, our formulation is based solely on distances to anchors, and no communication between the target nodes is assumed.

Finally, the Euclidean Distance Geometry (EDG) problem, which focuses on estimating the configuration of points from a subset of pairwise distances\citep{dokmanic2015euclidean,liberti2014euclidean,tasissa2018exact} has been applied to robust sensor localization \citep{biswas2006semidefinite}. For example, the semidefinite programming approach in \citep{biswas2006semidefinite} addresses robust sensor localization by using distances between sensors within radio range. In contrast, our approach relies solely on anchor-target distances and is supported by a theoretical analysis grounded in compressive sensing, which precisely characterizes when the target positions can be recovered from corrupted measurements.

\section{Motivation}
\label{sec:motivation}

\begin{figure}[h!]
	\centering
	\includegraphics[width=0.5\textwidth]{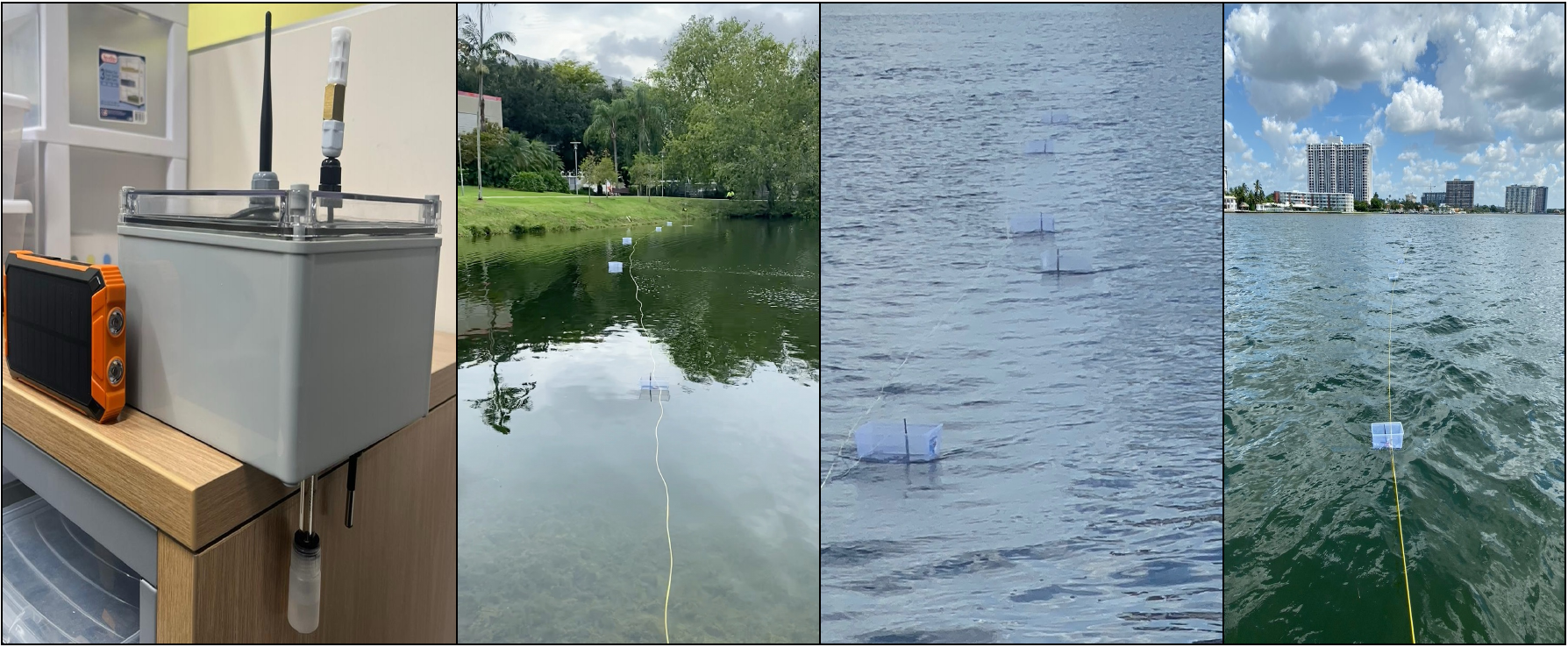}
	\caption{Deployment of low-power wireless sensor networks on the surface of different water bodies in Miami, Florida.}
	\label{fig:deployment}
\end{figure}

\begin{figure}[h!]
	\centering
	\includegraphics[width=0.45\textwidth]{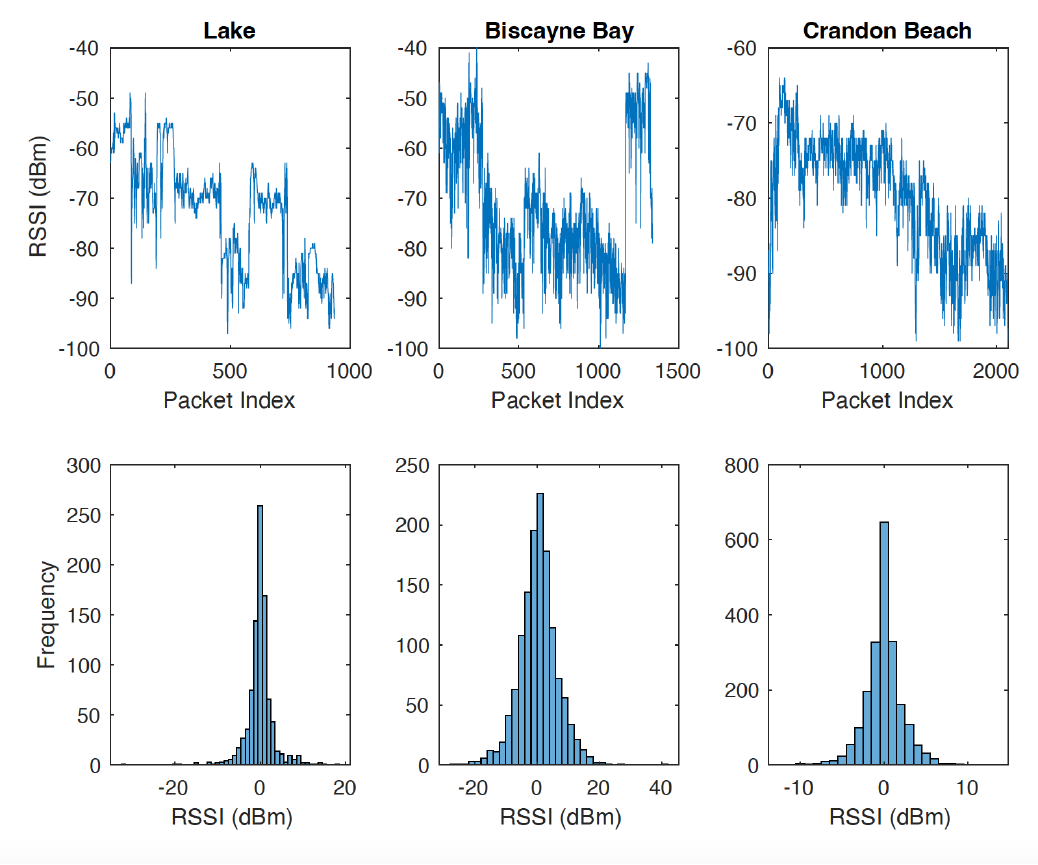}
	\caption{Link quality fluctuation as a result of the motion of water. The histograms refer to the change in received power (RSSI). Radio: CC2538, IEEE 802.15.4 compliant, low power radio operating in the 2.4 GHz frequency band (Texas Instruments).}
	\label{fig:rssi}
\end{figure}

Our work was motivated by practical deployments we undertook on the surface of different water bodies in Miami, Florida, to monitor water quality (Fig. \ref{fig:deployment}). The sensor nodes experienced different types of motions (localized as well as translational), the effects of which on the quality of the wireless links the nodes established were considerable (Fig. \ref{fig:rssi}). Indeed, most of the nodes experienced packet loss ranging from 30 to 60\%. This implies that radio-frequency (RF) based localization approaches for these types of deployments should accommodate and deal with a significant amount of error and signal corruption.

\section{Technical background}
\label{sec:background}

\subsection{Notation}

In this section, we summarize the notation used throughout this manuscript. Boldface uppercase letters denote matrices (e.g., $\bA$), and boldface lowercase letters denote column vectors (e.g., $\bx$).  The transpose of a vector $\bx$ is denoted by $\bx^\top$,
and the transpose of a matrix $\bX$, by $\bX^\top$. The $j$-th entry of a vector $\bx$ is denoted by $x_{j}$, and the $(i,j)$-th entry of a matrix $\bX$ is denoted by $X_{ij}$. The $i$-th column of the matrix $\bA$ is denoted by $\ba_i$. The vector $\one$ refers to a column vector of ones, and $\zero$ refers to a column vector of zeros, with their sizes determined by the context. The identity matrix is denoted by $\bm{I}$. We use $\Vert \bx \Vert$ to denote the standard $\ell_2$ norm on $\real^{n}$, defined as $\Vert \bx \Vert = \sqrt{\sum_{i=1}^{n} x_i^2}$. $\Vert \bx \Vert_1$ denotes the $\ell_1$ norm on $\real^{n}$, defined as $||\bx||_1=\sum_{i=1}^{n} |x_i|$. $\Vert \bX \Vert_F$ denotes the Frobenius norm on $\real^{m \times n}$, defined as $||\bX||_F = \sqrt{\sum_{i=1}^{m}\sum_{j=1}^{n} X_{i,j}^2}$. Given a matrix $\bX\in \real^{m\times n}$, the $\ell_{1,2}$ norm is defined as $\vectornorm{\bX}_{1,2}= \sum_{i=1}^{n} \vectornorm{\bx_i}_2$. In other words, the $\ell_{1,2}$ norm is the sum of the column norms of the matrix $\bX$. $\Vert \bX \Vert_*$ denotes the nuclear norm of a matrix $\bX$, defined as the sum of its singular values. Given a matrix $\bX\in \real^{m \times n}$, $\bX = \bU\bm{\Sigma}\bV^T$ denotes the singular value decomposition of $\bX$, where $\bU$ and $\bV$ are orthonormal matrices and $\bm{\Sigma}$ is a rectangular diagonal matrix that contains the singular values of $\bX$ along its diagonal. 

\begin{table}[h!]
\centering
\caption{Summary of notation}
\begin{tabular}{ll}
\hline
\textbf{Notation} & \textbf{Description} \\
\hline
$\bA$ & Matrix  \\
$\bx$ & Column vector  \\
$\bx^\top$ & Transpose of vector $\bx$ \\
$\bX^\top$ & Transpose of matrix $\bX$ \\
$x_j$ & The $j$-th entry of vector $\bx$ \\
$X_{ij}$ & The $(i,j)$-th entry of matrix $\bX$ \\
$\bm{a}_i$ & The $i$-th column of matrix $\bA$ \\
$\bm{1}$ & Column vector of ones  \\
$\bm{0}$ & Column vector of zeros  \\
$\bm{I}$ & Identity matrix \\
$\|\bx\|$ & Standard $\ell_2$ norm: $\| \bx \| = \sqrt{\sum_{i=1}^{n} x_i^2}$ \\
$\|\bx\|_1$ & $\ell_1$ norm: $\| \bx \|_1 = \sum_{i=1}^{n} |x_i|$ \\
$\|\bX\|_F$ & Frobenius norm: $\| \bX \|_F = \sqrt{\sum_{i=1}^{m} \sum_{j=1}^{n} X_{ij}^2}$ \\
$\|\bX\|_*$ & Nuclear norm: $\| \bX \|_* = \sum_{i=1}^{r} \sigma_i(\bX)$. \\
$\|\bX\|_{1,2}$ & $\ell_{1,2}$ norm: $\| \bX \|_{1,2} = \sum_{i=1}^{n} \| \bx_i \|_2$ \\
$\bX = \bU\bm{\Sigma}\bV^T$ & Singular value decomposition of $\bX$.\\
\hline
\end{tabular}
\end{table}

\subsection{Trilateration}
\label{trilateration}

\begin{figure}[h!]
    \centering
    \includegraphics[width=0.5\linewidth]{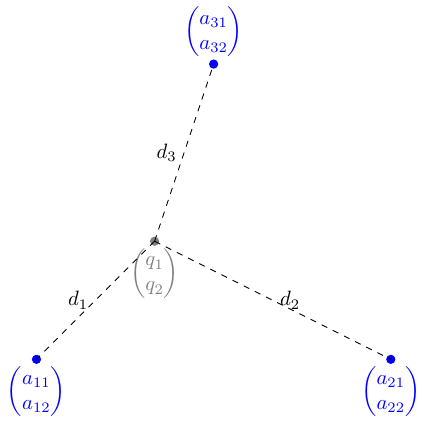}
    \caption{Illustration for trilateration in 2D: The aim is to estimate the position of the target node given the distances to three anchors.}
    \label{fig:trilateration_example}
\end{figure}

This section provides a brief background to trilateration. Given a target node in $\real^r$ and $r+1$ affinely independent anchors with known positions, the goal is to determine the target node's position based on its distances to the anchors. We assume the distances are exact and deem the last anchor as the central anchor node. We begin by illustrating the derivation of trilateration for the case of three anchors in a 2-dimensional space. Let $\bq=\begin{bmatrix}q_1\\q_2\end{bmatrix}$ represent the position of the target node, and let $\bx_1=\begin{bmatrix}a_{11}\\a_{12}\end{bmatrix}$, $\bx_2=\begin{bmatrix}a_{21}\\a_{22}\end{bmatrix}$, and $\bx_3=\begin{bmatrix}a_{31}\\a_{32}\end{bmatrix}$ denote the coordinates of the anchor nodes, respectively. Figure \ref{fig:trilateration_example} shows an illustration. 
We now express the squared distances from the target node to the three anchors as follows:
\begin{align*}
(q_1-a_{11})^2+(q_2-a_{12})^2 &= d_1^2\\
(q_1-a_{21})^2+(q_2-a_{22})^2 &= d_2^2\\
(q_1-a_{31})^2+(q_2-a_{32})^2 &= d_3^2.
\end{align*}
Subtracting the last equation from the first and second equations and simplifying, we obtain:
\begin{align*}
q_1(a_{11}-a_{31})+q_2(a_{12}-a_{32}) = c_1\\
q_1(a_{21}-a_{31})+q_2(a_{22}-a_{32}) = c_2,
\end{align*}
where 
\[
\begin{bmatrix}
    c_1\\c_2
\end{bmatrix}
= \frac{1}{2}\begin{bmatrix}
    d_3^2-d_1^2+\vectornorm{\bx_1}^2-\vectornorm{\bx_3}^2 \\
    d_3^2-d_2^2+\vectornorm{\bx_2}^2-\vectornorm{\bx_3}^2
\end{bmatrix}.
\]
Therefore, the trilateration problem is equivalent to solving the following linear system:
\[
\begin{bmatrix}
(\bx_1-\bx_3)^\top\\   
(\bx_2-\bx_3)^\top\\   
\end{bmatrix}
\begin{bmatrix}
    q_1\\q_2
\end{bmatrix} =\frac{1}{2}\begin{bmatrix}
    d_3^2-d_1^2+\vectornorm{\bx_1}^2-\vectornorm{\bx_3}^2\\
    d_3^2-d_2^2+\vectornorm{\bx_2}^2-\vectornorm{\bx_3}^2
\end{bmatrix}.
\]
It can be verified that the above linear system has a unique solution, if the three anchors are affinely independent i.e., the three anchors do not lie on a line in $\real^2$.\\

\noindent \textbf{Generalization}: We now consider the general problem of trilateration with $r+1$ anchor nodes in $\real^r$, where our goal is to estimate the position of a target node. We follow the same derivation as in the 2-dimensional case. Let $\bx_1, \dots, \bx_{r+1}$ denote the positions of the anchors in $\real^r$,
and $\bq\in \real^r$ denote the position of the target node. For ease of notation, we denote the last node, deemed as the central node, 
by $\bx_{c}$ i.e., $\bx_{r+1}=\bx_{c}$. The trilateration problem is equivalent to solving the following linear system:
\begin{equation*}
\begin{bmatrix}
   (\bx_1 - \bx_{c})^\top \\
   (\bx_2 - \bx_{c})^\top \\
   \vdots \\
   (\bx_r - \bx_{c})^\top
\end{bmatrix}
\begin{bmatrix}
    q_1 \\
    q_2 \\
    \vdots \\
    q_r
\end{bmatrix}
=\frac{1}{2}
\begin{bmatrix}
\bar{d}_{t,a_1}^2   + ||\bx_1||^2-||\bx_{c}||^2  \\
\bar{d}_{t,a_2}^2 + ||\bx_2||^2-||\bx_{c}||^2  \\
\vdots \\
\bar{d}_{t,a_r}^2  + ||\bx_r||^2-||\bx_{c}||^2
\end{bmatrix},
\end{equation*}
where $\bar{d}_{t,a_i}^2=\left(d_{t,a_{c}}^2-d_{t,a_1}^2\right)$, $\bm{q}\in \real^r$ denotes the unknown position of the target node, $d_{t,a_i}$ is the distance between the target node and the $i$-th anchor and $d_{t, a_c}$ is the distance between the target node and the the central node. If the anchors are affinely independent, the above system admits a unique solution, yielding the exact position of the target node. For further details on localization and its linear system formulation, we refer the reader to \citep{dargie2010fundamentals,wang2015linear,lichtenberg2024localization}.

\subsection{Compressive sensing}
\label{sec:compressed_sensing}

Underdetermined linear systems arise in many practical problems. For instance, in trilateration, determining the location of a target node in $\real^3$ given distances to only two anchors results in an underdetermined problem, as there are more unknowns than constraints (equations). For these systems, we consider the problem of solving $\bB \bq = \bc$, where $\bB \in \real^{d \times n}$ with $n \gg d$. Given that we have few constraints relative to the number of unknowns, such systems generally have infinitely many solutions, and obtaining a unique solution requires imposing additional structure on the unknown vector. One common approach is to assume sparsity, meaning the underlying vector has ``mostly" zero entries, though the locations of the nonzero entries are unknown. A common way to address this is through the following optimization program:
\begin{equation}\label{eq:l0_norm}
\minimize_{\bq \in \real^{n}} \,\, \vectornorm{\bq}_{0}\quad
 \subjectto \quad \bB\bq=\bc,
\end{equation}
where $\vectornorm{\cdot}_{0}$ denotes the $\ell_0$ norm of the vector in consideration, defined as the number of nonzero entries. A vector $\bq$ is $k$-sparse if it has at most $k$ nonzero entries i.e., $||\bq||_0\le k$. 
While minimizing the number of nonzero entries in the solution vector is a natural approach, it is challenging to solve directly. Since the location of the nonzero entries of $\bq$ are unknown, a naive approach would be to consider all possible locations where $\bq$ could be nonzero, starting with one nonzero entry, and then solving the equation $\bB\bq = \bc$ for each choice of nonzero locations, and selecting the solution with the fewest nonzero entries. However, even for a moderately sized vector $\bq$, this approach becomes computationally prohibitive. In fact, the complexity of this problem is known to be NP-hard \citep{natarajan1995sparse}. Examples of alternative tractable optimization algorithms for this are basis pursuit and greedy pursuit \citep{tropp2004greed,chen2001atomic,davis1997adaptive,pati1993orthogonal,tropp2007signal}. In basis pursuit, the idea is based on a convex relaxation of the $\ell_0$ norm, which is the $\ell_1$ norm. For this case, the optimization program is given by 
\begin{equation}\label{eq:l1_norm}
\minimize_{\bq \in \real^{n}} \,\, \vectornorm{\bq}_{1}\quad
 \subjectto \quad \bB\bq=\bc.
\end{equation}
Several fundamental questions have been studied in relation to \eqref{eq:l0_norm} and \eqref{eq:l1_norm}. The first is: under what conditions on the measurement matrix $\bB$ does the 
$\ell_0$ minimization program in \eqref{eq:l0_norm} admit a unique solution? The second is: what conditions ensure that solving the relaxed problem in \eqref{eq:l1_norm}
 yields the same solution to the original optimization problem \eqref{eq:l0_norm}? Various conditions on the measurement matrix address these questions \citep{candes2005decoding,foucart2013mathematical}. 
 
 In this manuscript, we focus on the mutual coherence condition which is easier to verify for a given candidate measurement matrix. The mutual coherence of a measurement matrix  $\bB$ is defined as
\begin{equation}\label{eq:coherence}  
\mu(\bB) = \underset{i\neq j}{\max}\,\, |\bb_i^T\bb_j|,
\end{equation}
where $\bb_i$ denotes the $i$-th column of $\bB$ which is assumed to be unit-norm. For a given measurement matrix $\bB$, we can normalize each column to be unit-norm before computing the coherence. We give two concrete examples of measurement matrices which respectively have high and low coherence. The first example is
\[
\bB = 
\begin{bmatrix*}
    1 & -1 & 1 & -1 \\
    -1 & 1 & 1 & -1 \\
    1 & 1 & -1 & 1 
\end{bmatrix*}.
\]
It can be easily verified that $\mu(\bB)=1$. The second example is
\[
\bB = 
\begin{bmatrix}
    1 & 1 &   1 & 1\\
    0 & 1 &   1 &-1\\
    0 & 1 &  -1 & 1
\end{bmatrix}.
\]
It is easy to verify that $\mu(\bB) = \frac{1}{\sqrt{3}}\approx 0.5574$. From these examples, we can infer that high coherence indicates vectors that are highly correlated, whereas low coherence suggests vectors that are uncorrelated and nearly orthogonal. The following technical result will be useful in this paper.

\begin{theorem}\label{thm:cs_theorem}[\citep{donoho2003optimally,gribonval2003sparse,tropp2004greed}.]
Consider $\bc = \bB \bq_*$, where $\bq_*$ is the underlying sparse vector. If $\vectornorm{\bq_*}_{0} < \frac{1}{2}\left(1 + \frac{1}{\mu(\bB)}\right)$, the minimization problem in \eqref{eq:l0_norm} has a unique solution, and the convex relaxation given in \eqref{eq:l1_norm} is tight- i.e., the optimal solution of \eqref{eq:l1_norm} is  the same as the optimal solution of \eqref{eq:l0_norm}.
\end{theorem}

\begin{remark}
 The theorem above describes the relationship between the coherence of $\bB$ and the number of nonzero entries in $\bq$. Notably, when coherence is low, we can accommodate more nonzero entries in $\bq$, while high coherence limits the number of permissible nonzero entries. In our context, as we will explain in upcoming sections, the number of nonzero entries corresponds to the number of outliers. Thus, measurement matrices with low coherence are desirable.
 \end{remark}

\begin{remark}
    The coherence of a measurement matrix captures the worst-case column correlation which could be pessimistic. Unlike the harder-to-verify Restricted Isometry Property (RIP), especially for deterministic matrices, coherence is easily computable while offering weaker recovery guarantees \citep{foucart2013mathematical}.
\end{remark}

\section{Problem Setup}
\label{sec:setup}

Consider a group of $m$ anchor nodes and $n$ target nodes in $\real^r$, where $r = 2$ or $3$. From each target node, we have distance measurements to the anchor nodes, and the positions of the anchors are known. However, distances between target nodes are not provided. This setup enables efficient algorithms that are based solely on anchor-target distances and eliminates the need for target-to-target communication, which is an ideal framework for resource-constrained or unreliable environments. When distance measurements are exact, under generic configuration of the anchor nodes and the central node, classic trilateration theory can be used to determine the exact positions of the target nodes. Formally, let $\bx_1,....,\bx_{m}$ denote the positions of the $m$ anchor nodes. Without loss of generality, we assume the central node is the $m$-th anchor node and is located at the origin. Using the same derivation as in trilateration (see Section \ref{trilateration}), the task of determining the target node's position can be formulated as the following linear system $\bX\bq=\bmm$:
\begin{equation}\label{eq:trilateration_system}
\begin{bmatrix}
   \dots\, \bx_1^\top\,\dots  \\
   \dots\,\bx_2^\top\, \dots\,   \\
    \vdots  \\
   \dots\,\bx_{m-1}^\top\, \dots\,
\end{bmatrix}
\begin{bmatrix}
    q_1 \\
    q_2 \\
    \vdots \\
    q_r
\end{bmatrix}
=
\begin{bmatrix}
\bar{d}_{t,a_1}^2 + ||\bx_1||^2 \\
\bar{d}_{t,a_2}^2 + ||\bx_2||^2 \\
\vdots \\
\bar{d}_{t,a_{m-1}}^2 + ||\bx_{m-1}||^2
\end{bmatrix},
\end{equation}
where $\bar{d}_{t,a_i}^2 = \frac{1}{2}\left(d_{t,a_{m}}^2-d_{t,a_i}^2\right)$ and $d_{t, a_i}$ is the distance between the target node and the $i$-th anchor. In compact form, this leads to the linear system $\bX\bq = \bmm$, where $\bX\in \real^{(m-1)\times r}$ and $\bmm\in \real^{m-1}$.

However, in practice, measurements are often corrupted due to environmental obstructions, sensor failures, or adverse conditions (ref. to Fig. \ref{fig:deployment}). For instance, in the setting we consider, target nodes may experience interference (e.g. heavy waves) that hinders communication with anchors, resulting in poor estimate of the true distances. In this manuscript, we study a setting where, among the $n$ target nodes, the distance measurements of $\alpha$ nodes are subject to severe corruption. Importantly, the identity of the corrupted nodes is also unknown in advance. Moreover, for a corrupted node, not all its distance measurements to anchors are necessarily affected and only a subset may be severely corrupted. This assumption ensures the possibility of robustly estimating the positions of the corrupted nodes. Our main assumptions are:
\begin{enumerate}
    \item 
 A random subset of the target nodes is prone to significant outliers in their distance measurements.
\item For each corrupted node, a random subset of its distances to anchors is severely corrupted.
\end{enumerate}
The problem is divided into two parts:
\begin{enumerate}
\item \textbf{Identification}: Identify the corrupted target nodes.
\item \textbf{Estimation}: Robustly estimate the locations of the corrupted nodes while minimizing the impact of the corrupted measurements.
\end{enumerate}
Before proceeding, we emphasize the importance of anchor and target node configurations for the identification problem to be well-posed. Consider a corruption model where a measured distance $d_{i,j}$ is perturbed to $d_{i,j}(1+\epsilon \cdot \mathcal{Z})$, with $\mathcal{Z}$ representing a random variable e.g., standard Gaussian noise (ref. to the histograms in Fig. \ref{fig:rssi}). This model introduces a relative error of approximately $\epsilon$, where the perturbation scales with the true distance. Consequently, small distances are less affected, while even slight corruption significantly impacts larger distances. Given this, under this corruption model, identifying corrupted nodes based solely on distances between targets and anchors is unreliable, particularly for corrupted nodes near anchors. To address this and ensure well-posedness, we assume a two-zone configuration: a near zone with target nodes close to anchors and a far zone with target nodes farther away. The corruption model selectively targets nodes in the far zone. This assumption is not only mathematically necessary but also practical, as nodes in the far zone are more prone to measurement errors.

\section{Robust estimation of the position of target nodes}
\label{sec:estimation}

In this section, we assume that the corrupted target nodes have been identified and set the focus on the robust estimation of their positions. The next section addresses the identification problem. While, in practice, identification must occur first, we discuss robust estimation first because some of the formulations developed for robust estimation can also be leveraged for identification. For clarity and pedagogical reasons, we present robust estimation before identification. For the robust position estimation problem, we assume that for each target node, the distances to a random subset of $k$ out of the $m$ anchors are highly corrupted. For the remaining nodes, we assume the distances are noisy but not highly corrupted. Recall that the trilateration equation is given by $\bX\bq = \bmm$. When all distances are exact and $\bX$ has full column rank, this equation yields the exact position of the target node. In what follows, we explain why the corruption of the distance from a target node to a central node should be treated in a special manner. Unlike the distance to a non-central anchor, which affects only its respective entry in $\bmm$, the distance to the central node appears in every entry of $\bmm$. Thus, corruption of the distance to the central node propagates throughout $\bmm$, making it a particularly challenging case. To address this, our model will account for the distance to the central node being corrupted.

Let $\tilde{\bmm}$ denote the corrupted distance measurements. We decompose $\bmm$ as follows: $
\tilde{\bmm} = \bmm + \bs + c\bm{1}$, where $\bmm$ represents the clean distance measurements, $\bs$ is a sparse or approximately sparse vector, capturing corruption of distances to $k$ or $k-1$ anchors (nonzero entries correspond to corrupted anchors, while small entries model exact measurements or noise) and $c\bm{1}$ is a constant vector modeling the possible uniform corruption due to the central node, with $c$ representing the unknown magnitude of this corruption. Recall that we can relate the exact measurements $\bmm$ with $\bq$ using the trilateration equation: $\bX\bq=\bmm$. Therefore, our goal is to estimate $c$, $\bs$, and $\bq$ (the position of the target node) given the corrupted measurements $\tilde{\bmm}$. 

We now introduce an algorithm for estimating the position of corrupted target nodes, leveraging an alternative formulation of the equation  $ \bX\bq + \bs + c\bm{1} = \tilde{\bmm}$ in terms of $\bs$. A critical assumption in this approach is that the augmented system $[\bX \,\, \bm{1}]$ is full column rank. In other words, the constant vector $\bm{1}$ must not lie in the column space of the anchors position matrix $\bX$. In most practical sensor localization scenarios, particularly with general or random deployment, we expect this assumption to hold generically.

Our approach is to employ ideas from robust error correction \citep{candes2005error,candes2008highly}
to eliminate the unknowns $\bq$ and $c$, and formulate the problem in terms of the sparse or approximately sparse vector $\bs$. Assuming $m > r + 2$, we define a matrix $\bar{\bR} \in \real^{(m-r-2) \times (m-1)}$ such that  
$\bar{\bR} \bX = \bm{0}$ and $\bar{\bR} \bm{1} = \bm{0}$. Multiplying both sides of the equation $\bX\bq + \bs + c\bm{1} = \tilde{\bmm}$ by $\bar{\bR}$, we obtain $\bar{\bR} \bs = \bar{\bR} \tilde{\bmm}$. This results in a compressive sensing problem for $\bs$, which can be solved using $\ell_1$-minimization:
\begin{align}\label{eq:l1_min_s}
&\minimize_{\bs \in \mathbb{R}^{m-1}} \quad \|\bs\|_1\\
&\subjectto \quad \bar{\bR} \bs = \bar{\bR} \tilde{\bmm}. \nonumber
\end{align}
Let $\bs_*$ be the optimal solution obtained from \eqref{eq:l1_min_s}. To estimate the target position $\bq_*$, we solve the linear system  $ [\bX \,\, \bm{1}]\begin{bmatrix} \bq \\ c \end{bmatrix} = \tilde{\bmm} - \bs_*$.
Since $[\bX \,\, \bm{1}]$ is assumed to be full column rank, this system has a unique solution, allowing us to extract the target position $\bq_*$. The full algorithm for estimating the position of corrupted target nodes is summarized in Algorithm \ref{alg:position_estimation_2}.
\begin{algorithm}
\caption{Robust estimation of target node's position} \label{alg:position_estimation_2}
\begin{algorithmic}
\STATE
\STATE \textbf{Input}: Position matrix of the anchors $\bX$ and corrupted measurements $\tilde{\bmm}$.  
\STATE \textbf{Step 1}: Define $\tilde{\bX}=[\bX\,\,\bm{1}]$. Find an orthonormal basis for the null space of $\tilde{\bX}^T$, and let $\bar{\bR}$ be the matrix whose rows consist of these basis vectors. 
\STATE \textbf{Step 2}:  Solve \eqref{eq:l1_min_s} to obtain the optimal solution $\bs_*$. 
\STATE \textbf{Step 3}: Solve $[\bX\,\,\bm{1}]\begin{bmatrix}\bq\\c\end{bmatrix}=\tilde{\bmm}-\bs_*$ and obtain $\bq_*$. 
\STATE \textbf{Output}:  $\bq_*$.
\end{algorithmic}
\end{algorithm}

\subsection{Computational Complexity of Algorithm \ref{alg:position_estimation_2}}
\label{sec:computational_complexity_1}

In this section, we discuss the computational complexity of Algorithm \ref{alg:position_estimation_2} by detailing the cost of each step. The first step is to determine a basis for the null space of the appropriate matrices. This can be efficiently achieved using the singular value decomposition, at a computational cost of $O(mr^2 + r^3)$ \citep{trefethen2022numerical}. Next, the $\ell_1$ minimization problem in \eqref{eq:l1_min_s} can be solved using a standard method such as the primal-dual interior-point algorithm, with a computational cost of $O(m^3)$ \citep{monteiro1989interior,yang2013fast}.  
Finally, in Step 3 of the algorithm, a linear system needs to be solved to estimate the position of the target node. This can be done via QR factorization at a cost of $O(mr^2)$ \citep{trefethen2022numerical}.
Since the number of anchors is assumed to be much greater than $r$ (typically 2 or 3 in practical applications), the most computationally intensive step is solving the $\ell_1$ optimization problem. In the setting we consider where the number of anchor nodes is limited, the $O(m^3)$ cost is manageable. However, in applications where there are a sufficiently large number of anchors, fast $\ell_1$ minimization algorithms can be employed to solve the $\ell_1$ optimization program \citep{yang2013fast,beck2009fast,candes2005l1}.

\section{Identification of Severely Corrupted Nodes}
\label{sec:identification}

In this section, we address the problem of identifying severely corrupted nodes. We consider the matrix $\bF \in \real^{m \times n}$, which contains the squared pairwise distances between the anchors and target nodes. As discussed in Section \ref{sec:setup}, $\alpha$ of the columns in this matrix are highly corrupted, and our goal is to identify which ones. We build upon the mathematical formulation introduced in the previous section. Recall that for a given target node, the corrupted measurements can be decomposed as: $\tilde{\bmm} = \bX\bq + \bs + c\bm{1}$. For $n$ target nodes, this extends to: $\tilde{\bmm}_i = \bX\bq_i + \bs_i + c_i\bm{1}$, where the index $i$ corresponds to the $i$-th target node.
Assuming $m > r + 2$ (i.e., at least 5 anchors in 2D and 6 anchors in 3D), there exists a matrix $\bar{\bR} \in \mathbb{R}^{(m-r-2) \times (m-1)}$ such that: $\bar{\bR}\bX = \bm{0} \quad \text{and} \quad \bar{\bR}\bm{1} = \bm{0}$.
Multiplying both sides of the equation $\tilde{\bmm}_i = \bX\bq_i + \bs_i + c_i\bm{1}$ by $\bar{\bR}$, we obtain:
$\bar{\bR}\bs_i = \bar{\bR}\tilde{\bmm}_i$. Define $\bS = [\bs_1, \bs_2, \dots, \bs_n]$ as the structured outlier matrix, where each $\bs_i$ is its $i$-th column. Similarly, let $\tilde{\bMM} = [\tilde{\bmm}_1, \dots, \tilde{\bmm}_n]$, where each $\tilde{\bmm}_i \in \mathbb{R}^{m-1}$. This allows us to compactly write the $n$ independent linear problems as: $\bar{\bR}\bS = \bar{\bR}\tilde{\bMM}$. 

To promote structured (block) sparsity in $\bS$, we employ the $\ell_{1,2}$-norm, which is defined as the sum of the column norms of a matrix \citep{eldar2010block}. The corresponding optimization problem is formulated as:
\begin{equation}\label{eq:l12_minimization}    
\minimize_{\bS \in \mathbb{R}^{(m-1)\times n}} \quad  \|\bS\|_{1,2} \quad  \text{subject to} \quad \bar{\bR}\bS = \bar{\bR}\tilde{\bMM}.
\end{equation}
Here, $\|\bS\|_{1,2}$ serves as the structured sparsity-inducing term. After solving the optimization problem in \eqref{eq:l12_minimization}, we obtain the optimal solution $\bS_*$. To identify the highly corrupted nodes, we compute the column norms of $\bS_*$ and select the indices corresponding to the $\alpha$ largest norms. These indices are designated as the highly corrupted nodes. The algorithm for identifying the highly corrupted nodes is summarized in Algorithm \ref{alg:identification}.

\begin{algorithm}[H]
\caption{Identify highly corrupted nodes} \label{alg:identification}
\begin{algorithmic}
\STATE 
\STATE \textbf{Input}: Position matrix of the anchors $\bX$ and corrupted measurements $\tilde{\bmm}_i$
corresponding to each target node.  
\STATE \textbf{Step 1}: Define $\tilde{\bX}=[\bX\,\,\bm{1}]$. Find an orthonormal basis for the null space of $\tilde{\bX}^T$, and let $\bar{\bR}$ be the matrix whose rows consist of these basis vectors. 
\STATE \textbf{Step 2}:  Solve \eqref{eq:l12_minimization} to obtain the optimal solution $\bS_*$. Compute the
column norms of $\bS$.  
\STATE \textbf{Step 3}:  Let $\{i_1,i_2,...,i_{\alpha}\} \subset \{1,...,n\}$ denote the indices corresponding to the $\alpha$ largest column norms of $\bS^*$.
\STATE \textbf{Output}:  $i_1,i_2,...,i_{\alpha}$ are the highly corrupted nodes. 
\end{algorithmic}
\end{algorithm}

\subsection{Computational Complexity of Algorithm \ref{alg:identification}}

The first step of Algorithm \ref{alg:identification} costs $O(mr^2 + r^3)$ (see Section \ref{sec:computational_complexity_1}). In the second step, the problem decouples into solving $n$ independent minimum-norm least squares problems, which can be efficiently computed using QR decomposition at a cost of $O(nm^3)$. In Step 3, computing the column norms requires $O(nm)$, and sorting them to identify the $\alpha$ largest incurs a cost of $O(n \log n)$. Thus, the dominant cost is from Step 2. Note that, the number of target nodes generally far exceeds the number of anchors. With that, that the overall cost of the identification algorithm is favorable as it scales linearly with the number of target nodes.

\section{Theoretical analysis}
\label{sec:analysis}

In this section, our primary goal is to study Algorithms \ref{alg:position_estimation_2} and \ref{alg:identification} and to determine the conditions under which exact identification of corrupted target nodes and precise estimation of their positions can be achieved. We begin with the analysis for the problem of estimating the position of a single corrupted target node. This choice is motivated by the fact that the analysis for this case serves as a fundamental step in our analysis of identification of  corrupted target nodes.

\subsection{Exact estimation of positions of corrupted target nodes}

Recall that for each corrupted target node, we assume that $k$ of its distances to anchor nodes are highly corrupted. For the following analysis, we assume that the remaining distances are exact. Moreover, under this model, the level of corruption can be arbitrary. The central question is whether it is possible to recover the true position of the corrupted target node. In \eqref{eq:l1_min_s}, we established a sparse recovery program to obtain an optimal solution $\bs_*$. Given $\bs_*$, we estimate the target position by solving the linear system:
$[\bX\,\,\bm{1}]\begin{bmatrix}\bq\\c\end{bmatrix}=\tilde{\bmm}-\bs_*$. See Algorithm~\ref{alg:position_estimation_2}, which summarizes these two steps. Our objective is to determine the conditions under which Algorithm~\ref{alg:position_estimation_2} successfully recovers the true target position $\bq_*$. To achieve this, we use theoretical guarantees from compressive sensing (see Theorem~\ref{thm:cs_theorem}), which specifically relies on the coherence of the measurement matrix.

Note that $\bs$ is a vector with at most $k$ nonzero entries, as only $k$ of the distance measurements to the anchors are corrupted. We now state the main result, which follows from applying Theorem~\ref{thm:cs_theorem} to \eqref{eq:l1_min_s}.
\begin{theorem}\label{thm:position_estimation_theorem}
Let $[\bX\,\,\bm{1}]$ be a full column rank matrix and  
$\bar{\bR}\bs_* = \bar{\bR} \tilde{\bmm},
$
where $\bs_*$ is the underlying $k$-sparse vector. If 
\begin{equation*}
k < \frac{1}{2}\left(1 + \frac{1}{\mu(\bar{\bR})}\right),
\end{equation*}
then the following minimization problem yields a unique solution $\bs_*$:
\begin{equation*}\label{eq:l1_robust}
\minimize_{\bs \in \real^{m-1}} \,\, \vectornorm{\bs}_{1} \quad
\subjectto \quad \bar{\bR}\bs=\bar{\bR}\tilde{\bmm}.
\end{equation*}
Consequently, the exact position of the target node $\bq_*$ can be recovered by solving
\begin{equation*} [\bX\,\,\bm{1}]\begin{bmatrix}\bq\\c\end{bmatrix}=\tilde{\bmm}-\bs_*.   
\end{equation*}
\end{theorem}
\begin{proof}
    The proof follows directly from applying Theorem \ref{thm:cs_theorem} and the fact that the full column rank of $[\bX\,\,\bm{1}]$ guarantees a unique target position $\bq_*$. 
\end{proof}
To accommodate a larger number of outliers $k$, it is desirable for the coherence of the augmented matrix $\bar{\bR}$ to be small. In a later section, we will discuss deployment configuration designed to minimize the coherence of $\bar{\bR}$. 
\subsection{Identifying Corrupted Target Nodes}

In this section, we study the problem of accurately identifying corrupted target nodes. We assume that $\alpha$ of the nodes are highly corrupted, with the level of the corruption being arbitrarily severe while the remaining nodes have exact measurements. Let $\bS_* \in \mathbb{R}^{(m-1) \times n}$ denote the underlying outlier matrix, where $\alpha$ of its columns are corrupted, and the remaining columns only have zero entries. As discussed in Section \ref{sec:identification}, we can conclude that $\bR \bS_* = \bR \tilde{\bMM}$. This equation carries an important
implication. If the $j$-th column of $\bS_*$ is a zero vector (indicating that the corresponding target node is not corrupted), then the $j$-th column of $\bR \tilde{\bMM}$  is a zero vector. Conversely, if the $j$-th column of $\bS_*$ is a dense vector (indicating corruption), then the corresponding column of $\bR \tilde{\bMM}$  is also dense. 
Therefore, to identify the $\alpha$ corrupted target nodes under this setup, it suffices to apply the matrix $\bR$ to the matrix of corrupted measurements $\tilde{\bMM}$, and identify the $\alpha$ non-zero columns. This identification does not need to solve the optimization program in \eqref{eq:l12_minimization}. The simplicity of this identification process is due to the assumption that the
distance measurements for uncorrupted target nodes are exact. Although our analysis focuses on this idealized scenario, our empirical experiments show that \eqref{eq:l12_minimization} performs well in identifying corrupted nodes even when there are moderate errors in the distance measurements for the uncorrupted nodes. A rigorous analysis of Algorithm 2 for the general setting remains an interesting direction for future research. We believe that employing configurations with target nodes placed in both near and far zones and concentrating the corruption on the nodes in the far zones, could facilitate such an analysis.

\section{Optimal deployment configuration}
\label{sec:config}

Theorem \ref{thm:position_estimation_theorem}  establishes the conditions for exact localization. Beyond the theoretical analysis, Algorithm \ref{alg:position_estimation_2} provides a practical solution for this task. A key component in the theorem is the coherence of the matrix $\bar{\bR}$. Since coherence and the number of corrupted measurements are inversely related, low-coherence measurement matrices are highly desirable. This leads to the question: what is the minimum coherence attainable by a generic matrix $\bG\in \real^{s\times t}$ with $t>s$, assuming unit-normed columns? The Welch bound \citep{welch1967lower} provides an answer:
\[
\mu \ge \sqrt{\frac{t-s}{s(t-1)}}.
\]
This bound is tight and achieved by equiangular tight frames (ETF) \citep{strohmer2003grassmannian}, though ETFs exist only for specific choices of $s$ and $t$ and require careful construction \citep{jasper2013kirkman,fickus2012steiner}. 

In this manuscript, we seek to optimize the deployment configuration, in the sense of minimizing the coherence of $\bar{\bR}$. However, directly optimizing $\bX$ for this objective does not yield a simple optimization program. Instead, we take an inverse approach: we construct $\bar{\bR}$ to achieve minimal coherence and then derive the corresponding anchor configuration from its structure. This approach is simpler and more direct, assuming some flexibility in choosing the deployment configuration. We detail its construction below.

The core idea is to find a matrix $\bG\in \real^{s\times t}$, with $t> s$ and rank $s$, that minimizes coherence, i.e., encourages its columns to be as orthonormal as possible. This leads to the optimization problem:
\begin{equation}\label{eq:coherence_minimize}
\minimize_{\bG \in \real^{s\times t}} \,\, 
\frac{1}{2}\|\bG^T\bG - \bm{I}\|_F^2.
\end{equation}
The next theorem establishes a simple solution to this problem.
\begin{theorem}
\label{thm:coherence_min_theorem}
The optimal full row rank solution $\bG_*$ to the optimization program \eqref{eq:coherence_minimize} is given by $\bG_* =\bm{U}\bm{V}^T$, where $\bm{U}\in \real^{s\times s}$ is any orthonormal matrix and $\bm{V}\in \real^{t\times s}$ is any matrix with orthonormal columns. 
\end{theorem}
Based on the above theorem, Algorithm \ref{alg:optimize_coherence} discusses a construction of the matrices $\bm{U}$ and $\bm{V}$ such that $\bar{\bR} \in \real^{(m-r-2)\times (m-2)}$ has the minimal coherence while satisfying $\bar{\bR}\bm{1}=\bm{0}$. 
\begin{algorithm}
\caption{Finding a low coherence matrix $\bar{\bR}$ such that $\bar{\bR} \bm{1} = \bm{0}$} 
\label{alg:optimize_coherence}
\begin{algorithmic}[1]
\STATE \textbf{Step 1}: Construct an orthonormal basis for the null space of $\bm{1}^T$. Let $\tilde{\bm{V}} \in \mathbb{R}^{(m-1) \times (m-2)}$ be the matrix whose columns contain these basis vectors.

\STATE \textbf{Step 2}: Perform the QR decomposition of a random matrix $\bm{A}$, and set $\bm{U} = \bm{Q}$.

\STATE \textbf{Step 3}: Define $\bm{V} \in \mathbb{R}^{(m-1) \times (m-r-2)}$ as the matrix whose $m-r-2$ columns are the first $m-r-2$ columns of $\tilde{\bm{V}}$.

\STATE \textbf{Step 4}: Compute $\bar{\bR}$ as: $\bar{\bR} = \bm{U} \bm{V}^T$.

\STATE \textbf{Output}: The low-coherence matrix $\bar{\bR}$.
\end{algorithmic}
\end{algorithm}
\noindent To determine the optimal anchor-position matrix $\bX_*$, it suffices to find a basis for the null space of $\bar{\bR}$, and set the columns of $\bX_*$ to be the first $r$ basis vectors. 

\section{Numerical Experiments}
\label{sec:experiments}

In this section, we present numerical results to evaluate the performance of the proposed algorithms. All algorithms were implemented in MATLAB R2022a, and the experiments were conducted on a Lenovo ThinkPad X1 Carbon equipped with an Intel Core i5-1135G7 processor (4 cores, 8 logical processors) and 16 GB of RAM. To solve Algorithm 2,
 we use CVX \citep{cvx,gb08} Specifically, we use the Mosek 10.2.1 solver \citep{mosek}. The code will be publicly available in a repository upon acceptance of this paper.   

\begin{figure}
    \centering    \includegraphics[width=0.6\linewidth]{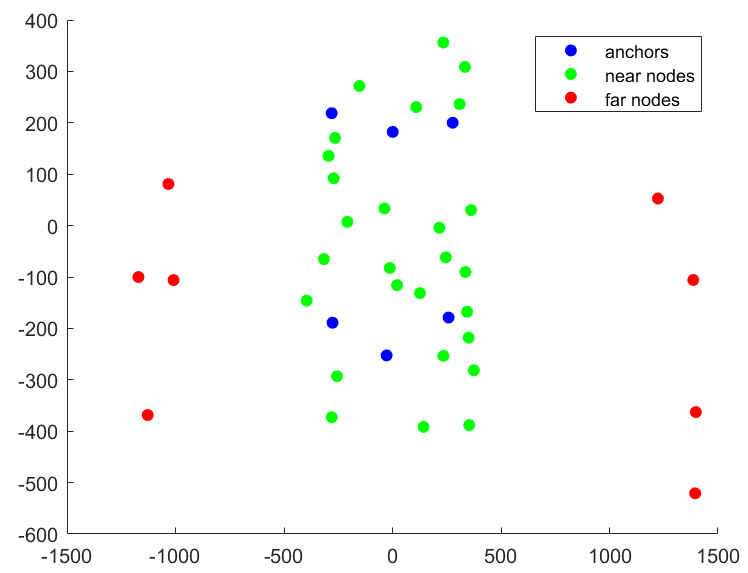}
    \caption{Visualization of the region for the anchor nodes and target nodes. }
    \label{fig:region_visualization}
\end{figure}

\begin{figure}
    \centering    \includegraphics[width=0.6\linewidth]{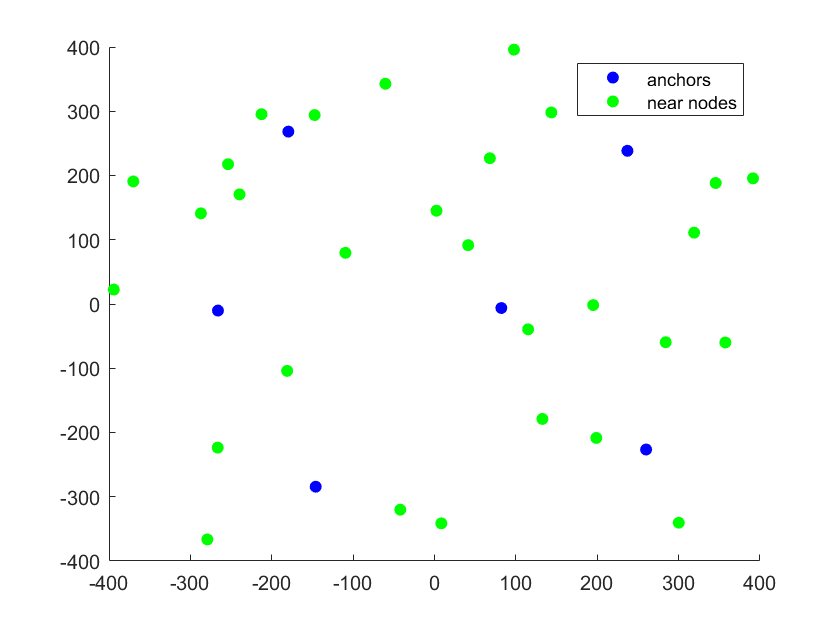}
    \caption{Visualization of the region for the anchor nodes and near target nodes. The anchors are spatially distributed. }
    \label{fig:region_nearby}
\end{figure}

\noindent \textbf{Region setup and node placement}: For all numerical experiments, we first generate $1,000$ points uniformly distributed within a region where both the 
$x$- and $y$-coordinates range from $[-400, 400]$. We then apply K-means clustering to partition these points into 
$m$ clusters. The cluster centers obtained from K-means serve as the positions of the anchors, with the final anchor designated as the central node. Next, we generate the target nodes as follows. For the near target nodes,
we randomly sample $50$ points uniformly within the same region. To ensure that nodes are well-spaced, we exclude nodes that are too close to anchors or to each other, using a minimum distance threshold of $60$. The resulting set of target nodes constitute the near target nodes. For the far target nodes, we generate additional target nodes in two separate
regions. The first region has its $x$-coordinates in $[-1200,-1000]$, and its $y$-coordinates in $[-600,400]$.
The second region has its $x$-coordinates in $[1200,1400]$ and its $y$-coordinates in $[-600,400]$.   In each region, we sample 50 uniformly distributed target nodes and enforce a minimum separation distance of $60$ between the nodes. The resulting nodes constitute the far target nodes. Figure \ref{fig:region_visualization} illustrates a sample instance of the region, showing the anchor nodes, the near target nodes, and far target nodes. Figure \ref{fig:region_nearby} illustrates another instance of the anchor nodes and the near target nodes. Note that while the anchors might look clustered in Figure \ref{fig:region_visualization} due to the scale, they are spatially distributed in the near zone.   

\noindent \textbf{Corruption model}: From the set of far target nodes, we randomly sample $\alpha$ nodes, marking them as highly corrupted nodes. The remaining nodes are designated as normal target nodes. The corruption model modifies the distances between each target node and its anchors as follows. For normal target nodes, distances are corrupted as: $d_{i,j}^2 = (1+\mathcal{U}(a,b)) \cdot d_{i,j}^2$ where $\mathcal{U}(a,b)$ is a uniform random variable in the interval $[a, b]$. For the highly corrupted nodes, a random subset of the $k$ distances, are corrupted following the same form but with a different perturbation range: $d_{i,j}^2 = (1+\mathcal{U}(c,d)) \cdot d_{i,j}^2$. We note that the remaining set of distances are corrupted with the same corruption model as the normal target nodes.

\noindent \textbf{Performance metrics}: We evaluate the algorithm’s performance based on two key tasks. The first
is the identification of corrupted target nodes, and the second is the robust estimation of target positions. For these, we use the following metrics:
\begin{enumerate} [leftmargin=*]
    \item Identification accuracy (IA): This metric measures the percentage of target nodes correctly identified as corrupted.

\item Mean relative error (MRE): The relative error of the estimated position is defined as 
\[
  \frac{\| \bq_* - \bq_{\text{est}} \|}{\| \bq_* \|},
  \]
  where $\bq_*$ is the true position of the target node, and $\bq_{\text{est}}$ is the estimated position. The mean relative error is the average of the relative error of the estimated positions over all $\alpha$ corrupted nodes.

\item Mean-square error of the positions (MSP): The position error is defined as  
  \[
  \|\bq^* - \bq_{\text{est}} \|^2,
  \]
  representing the squared distance between the true and estimated positions. The mean-square error of the positions is the average of the mean-square error across all $\alpha$ corrupted nodes.

\item Mean-square error of the distances (MSD): It is defined as:
 \[
  \frac{1}{\alpha m} \sum_{i=1}^{\alpha} \sum_{j=1}^{m} (d_{i,j} - \hat{d}_{i,j})^2,
  \]
  where $\hat{d}_{i,j}$ denotes the corrupted distance between $i$-th target node and the $j$-th anchor node. This metric quantifies the error inherent in the provided distances. This metric is useful as it provides the minimum MSP that can be attained by an ideal localization algorithm \citep{moore2004robust,savvides2003error}. 

\item Mean anchor distance ratio (MADR): For each target node, comute the following quantity:  
\[
\frac{\sum_{j=1}^{m} ||\bq_{\text{est}}-\ba_j||}{\sum_{j=1}^{m} ||\bq_*-\ba_j|2}. 
\]
The mean anchor distance ratio is the average of the above quantity over the $\alpha$ target nodes. 

\end{enumerate}

\noindent \textbf{Baseline algorithms for the identification problem:} 
We consider two baseline algorithms for identifying corrupted nodes. In our setting, a random subset of $\alpha$ columns of $\bF$ is severely corrupted, while the remaining columns are affected by noise. Let $\tilde{\bF}$ denote the corrupted matrix. A common approach for identifying corrupted nodes is based on \emph{structured robust PCA}, which solves the following optimization problem:
\begin{equation}
\label{eq:nuc_l1l2}    
\min_{\bF, \bS} \,\, \|\bF\|_* + \lambda \|\bS\|_{1,2} \quad \text{subject to} \quad \tilde{\bF} = \bF + \bS,
\end{equation}
where $\|\bF\|_*$ is the nuclear norm of $\bF$, which promotes low-rankness, $\|\bS\|_{1,2}$ is a structured sparsity-inducing term, and $\lambda$ is a regularization parameter that balances the two terms. Larger values of $\lambda$ promote sparser solutions for $\bS$, whereas smaller values allow for denser solutions. 

After solving the optimization problem in \eqref{eq:nuc_l1l2}, we obtain the optimal solutions $\bF^*$ and $\bS^*$. To identify the highly corrupted nodes, we first compute the column norms of $\bS^*$. The indices corresponding to the $\alpha$ largest norms of $\bS^*$ are then designated as the highly corrupted nodes. The choice of the regularization parameter $\lambda$ is crucial in balancing the trade-off between the low-rank structure of $\bF$ and the structured sparsity of $\bS$ in \eqref{eq:nuc_l1l2}. For our experiments, we consider the values $\lambda \in \{1, 2, 5, 10, 20\}$. After solving \eqref{eq:nuc_l1l2} for each $\lambda$, we obtain $5$ sets with $\alpha$ indices each. We then take the union of these sets and designate the corrupted nodes as the top $\alpha$ indices that appear most frequently. We refer to this baseline algorithm as \emph{SRPCA}. The second baseline algorithm is a \emph{naive method}, which simply selects the $\alpha$ columns with the highest column norms in the corrupted squared distance matrix of the target nodes and anchors. For both algorithms, accuracy is defined as the proportion of correctly identified corrupted nodes.

\subsection{Experiment 1: Idealized case with no corruption in normal target nodes}

In this experiment, we consider an idealized setting where the distances from the normal target nodes remain uncorrupted, while the corrupted target nodes are susceptible to distance corruptions. The primary objective is to test the algorithm’s performance under these ideal conditions: Can it achieve perfect identification and exact estimation of target node positions? We set the parameters as follows: $k = 3$ and $\alpha = 4$. Since there is no corruption in the normal nodes, $a = b = 0$. The corruption parameters for highly corrupted nodes
are $c = 0.2$ and $d = 0.25$. For each target node, a random subset of $k$ distances is corrupted, following the distribution $\mathcal{U}(c,d)$. This setup assumes exact distance information between target nodes and anchors, except for the highly corrupted nodes. For these nodes, all but $k$ distances remain exact, with the selected $k$ distances subjected to uniform noise based on the specified parameters. The relative error in the squared distances for the selected $k$ distances from corrupted nodes to anchors falls within the $20\%-25\%$ range. Table \ref{tab:accuracy_comparison_exact} compares the identification accuracy of the proposed algorithm with the structured robust PCA approach and the naive approach.  Table \ref{tab:estimation_metrics_exact} presents the performance of robust estimation of the corrupted target nodes, evaluated across the four performance metrics.

\begin{table}[ht]
 \caption{Comparison of Identification Accuracy (IA) between the proposed algorithm, SRPCA, and naive method for varying values of $m$ in Experiment 1. Reported values are averages of $50$ independent trials.}
    \label{tab:accuracy_comparison_exact}
    \centering
    \begin{tabular}{c|c c c}
        \hline
        $m$ & Ours   & SRPCA   & Naive   \\
        
         & accuracy (\%) &  accuracy (\%) &  accuracy (\%) \\
        \hline
        6  & 100 & 83.00 & 38.50 \\
        9  & 100 & 95.00 & 49.50 \\
        12 & 100 & 91.00 & 37.50\\
        15 & 100 & 85.50 & 40.50\\
        \hline
    \end{tabular}
   
\end{table}

\begin{table}[ht]
 \caption{Mean relative error (MRE), mean-square error of positions (MSP), mean-square error of distances (MSD), and Mean anchor distance ratio (MADR) for different values of $m$ for different values of $m$ for Experiment 1. Reported values are averages of $50$ independent trials.}
    \label{tab:estimation_metrics_exact}
    \centering
    \begin{tabular}{c|c c c c}
        \hline
        $m$ & MRE & MSP & MSD & MADR \\
        \hline
        6  & 0.329 & $2.192 \times 10^5$ & $1.051 \times 10^4$ & 1.132 \\
        9  & $6.080\times10^{-2}$  & $3.430 \times 10^4$ & $7.311 \times 10^3$ & 0.992 \\
        12 & $4.200\times 10^{-3}$ & $2.337 \times 10^3$ & $5.234 \times 10^3$ & 1.000 \\
        15 & $1.558 \times 10^{-12}$ & $3.345 \times 10^{-17}$ & $4.324 \times 10^3$ & 1.000 \\
        \hline
    \end{tabular}
   
\end{table}

\subsection{Experiment 2: Realistic case with corruption in normal target nodes}

\begin{figure}
    \centering
    \includegraphics[width=0.68\linewidth]{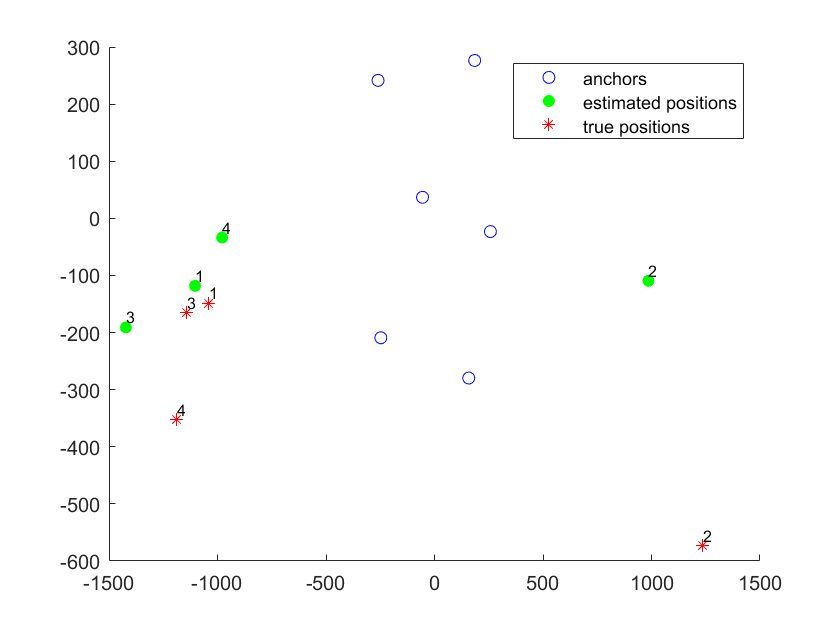}
    \caption{Comparison of the ground position and estimated position for the corrupted nodes for Experiment $2$. The number of anchors is $6$.}
    \label{fig:noisy_6_vis}
\end{figure}

\begin{figure}
    \centering
    \includegraphics[width=0.68\linewidth]{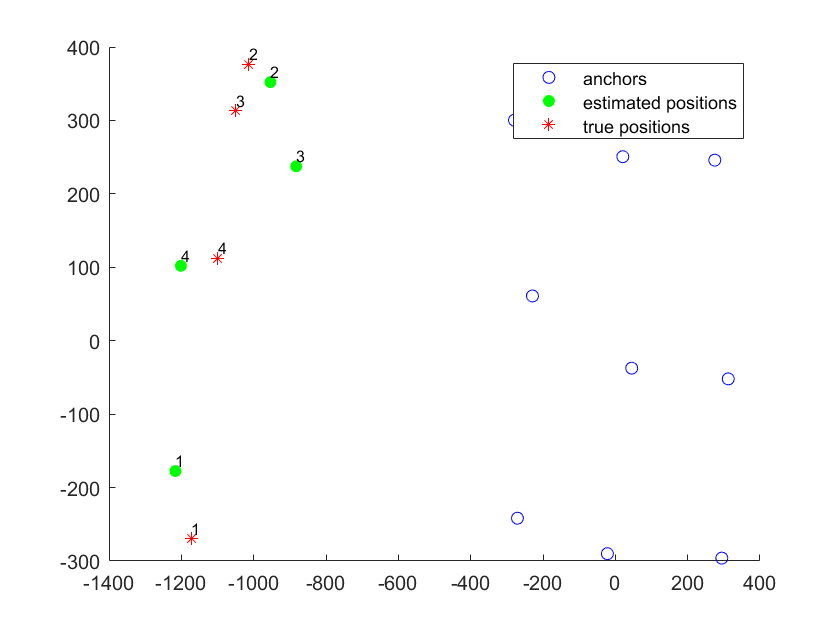}
    \caption{Comparison of the ground position and estimated position for the corrupted nodes for Experiment $2$. The number of anchors is $9$.}
    \label{fig:noisy_9_vis}
\end{figure}

In this experiment, we consider a more realistic setting where distance corruption affects both normal and corrupted nodes. The goal is to evaluate the algorithm's performance when all nodes experience some level of corruption.  
We set the parameters as follows: $k = 3$ and $\alpha = 4$. For the normal nodes,
the corruption parameters are $a = 0$ and $b= 0.05$. For the highly corrupted nodes,
the corruption parameters are $c = 0.15$ and $d = 0.2$. For each corrupted target node, a random subset of $k$ distances is corrupted according to the uniform distribution $\mathcal{U}(c,d)$. The relative error in the squared distances for the selected  $k$ distances from corrupted nodes to anchors falls within the $15\%-20\%$ range, while for all other distances, the relative error remains in the $0-5\%$ range. Table \ref{tab:accuracy_comparison_corrupted} compares the identification accuracy of the proposed algorithm with a naive approach. Table \ref{tab:estimation_metrics_corrupted} presents the performance of robust estimation of the corrupted target nodes, evaluated using four key performance metrics. Furthermore, Figures \ref{fig:noisy_6_vis} and \ref{fig:noisy_9_vis} show the comparison between the ground-truth positions of the corrupted target nodes and their estimated positions for $m=6$ and $m=9$ respectively. 

\begin{table}[h!]
 \caption{Comparison of Identification Accuracy (IA) between the proposed algorithm, SRPCA, and naive method for varying values of $m$ in Experiment 2.  Reported values are averages of $50$ independent trials.}
    \label{tab:accuracy_comparison_corrupted}
    \centering
   \begin{tabular}{c|c c c}
        \hline
        $m$ & Ours   & SRPCA.   & Naive   \\
        
         & accuracy (\%) &  accuracy (\%) &  accuracy (\%) \\
        \hline
        6  & 88.50 & 72.00 & 49.50 \\
        9  & 98.50 & 70.00 & 48.00 \\
        12 & 99.50 & 70.50 & 37.50\\
        15 & 97.50 & 69.50 & 42.00\\
        \hline
    \end{tabular}
   
\end{table}

\begin{table}[h!]
 \caption{Mean relative error (MRE), mean-square error of positions (MSP), mean-square error of distances (MSD), and Mean anchor distance ratio (MADR) for different values of $m$ for Experiment 2. Reported values are averages of $50$ independent trials.}
    \label{tab:estimation_metrics_corrupted}
    \centering
    \begin{tabular}{c|c c c c}
        \hline
        $m$ & MRE & MSP & MSD & MADR \\
        \hline
        6  & 0.272 & $1.444 \times 10^5$ & $6.611 \times 10^3$ & 1.231 \\
        9  & 0.120 & $3.710 \times 10^4$ & $4.472 \times 10^3$ & 1.048 \\
        12 & $6.77\times 10^{-2}$ & $1.020 \times 10^4$ & $3.421 \times 10^3$ & 0.947 \\
        15 & $5.40\times 10^{-2}$ & $6.225 \times 10^{3}$ & $2.758 \times 10^3$ & 0.988 \\
        \hline
    \end{tabular}
   
\end{table}

\subsection{Comparison with Existing Algorithm}

In this section, we compare our algorithm with the SR-Hybrid algorithm proposed in \citep{zaeemzadeh2017robust}.
We use the implementation of the algorithm in \citep{zaeemzadeh2024robustlocalization}, and use the default parameters. We select this algorithm for comparison because its setup is similar to ours, as both are based on the distances between target nodes and anchors. The experimental setup, except the corruption model, remains unchanged from previous sections. Specifically, the region configuration, as well as the placement of anchors and target nodes, is as described in Section \ref{sec:experiments}. To ensure a fair comparison with \citep{zaeemzadeh2017robust}, we adopt a similar corruption model for the distances. In \citep{zaeemzadeh2017robust}, each distance $d_{i,j}$ is corrupted as follows:  
$\bar{d}_{i,j} = d_{i,j} + \nu$, where $\nu$ is assumed to be independent and identically distributed noise term defined as $\nu = (1-\beta) \mathcal{N}(0,\sigma^2) + \beta \mathcal{U}(a,b)$, with $\mathcal{N}(0,\sigma^2)$ representing standard Gaussian noise with mean $0$ and variance $\sigma^2$, and $\mathcal{U}(a,b)$ denoting uniform random noise in the interval $[a,b]$. We emphasize that our general model accounts for multiplicative noise in squared distances, which is significantly more challenging. However, we adopt this simpler corruption model to ensure consistency with the approach used in \citep{zaeemzadeh2017robust}. For our experiments, we set $\beta = 0.75$, $\sigma = 0.01$, $a = -100$, and $b = 100$. Table \ref{tab:position_comparison_baseline} shows the results of the experiments, evaluated across the four performance metrics.

\begin{table}[h]
    \caption{Mean relative error (MRE), mean-square error of positions (MSP), mean-square error of distances (MSD), and Mean anchor distance ratio (MADR) for different values of $m$. Reported values are averages of $50$ independent trials for two different methods. For each $m$, the first row reports the results of Algorithm \ref{alg:position_estimation_2} while second row reports the results of the SRHybrid algorithm in \citep{zaeemzadeh2017robust}.}
    \label{tab:position_comparison_baseline}
    \centering
    \begin{tabular}{c|c|c c c }
        \hline
        $m$ &  MRE & MSP & MSD & MADR \\
        \hline
        \multirow{2}{*}{6}  &  0.134 & $3.391 \times 10^4$ & $1.823 \times 10^3$ & 0.944 \\
                            &  0.248 & $1.053 \times 10^5$ & $1.823 \times 10^3$ & 0.865 \\
        \hline
        \multirow{2}{*}{9}  &  0.109 & $2.321 \times 10^4$ & $1.771 \times 10^3$ & 0.994 \\
                            &  0.211 & $8.195 \times 10^4$ & $1.771 \times 10^3$ & 1.025 \\
        \hline
        \multirow{2}{*}{12} &  $0.105$ & $2.090 \times 10^4$ & $1.880 \times 10^3$ & 1.002 \\
                            &  $0.227$ & $9.378 \times 10^4$ & $1.880 \times 10^3$ & 0.967 \\
        \hline
        \multirow{2}{*}{15} &  $8.93\times 10^{-2}$ & $1.516 \times 10^{4}$ & $1.856 \times 10^3$ & 1.039 \\
                            &  $0.249$ & $1.067 \times 10^{5}$ & $1.856 \times 10^3$ & 1.191 \\
        \hline
    \end{tabular}
\end{table}

\section{Discussion}
\label{sec:discussion}

Our numerical results for Experiment 1 demonstrate that the proposed algorithm successfully identifies all corrupted nodes (see Table \ref{tab:accuracy_comparison_exact}), and we achieve favorable performance metrics with a configuration of $9$ anchors (see Table \ref{tab:estimation_metrics_exact}). In the more challenging scenario of Experiment 2, our identification accuracy outperforms the competitive methods considered (see Table \ref{tab:accuracy_comparison_corrupted}). Moreover, across the four performance metrics, good results (defined as having MSP/MSD below 5) are obtained when $m=12$ (see Table \ref{tab:estimation_metrics_corrupted}). Finally, comparing our method to the algorithm proposed in \citep{zaeemzadeh2017robust} suggests that our approach can handle different noise models, outperforming the competitive method for all numbers of anchors. Therefore, in the motivating problem we consider, if there are sufficiently many nodes near the anchors and relatively few nodes away, our results suggest that about $10\%$ of the nodes could serve as anchors for the tasks of identifying corrupted nodes and robustly estimating positions.

The strength of our proposed model is its reliance solely on anchor-target distances, which is ideal in resource-constrained settings. Both the identification and robust estimation problems are simple and come with theoretical guarantees under certain conditions. Furthermore, by using stability as a criterion, our method offers a way to construct optimal deployment configurations. Additionally, in the event of a sensor failure or adversarial attack, the identification process alone can highlight susceptible nodes.
One limitation of the model is that it assumes susceptible nodes are located further from the anchor nodes. In our mathematical formulation, this assumption is necessary, as large perturbations on small distances might otherwise be overlooked compared to relatively small perturbations on large distances. In practice, even nodes near the anchors could be highly corrupted, and incorporating this possibility into the formulation would be beneficial. Our approach leverages sparsity to develop the proposed optimization algorithms; however, capturing the mathematical characterizations of the structured noise inherent in localization problems could further broaden the applicability of our method.

\section{Conclusion}
\label{sec:conclusion}

In this manuscript, we address the localization problem for networks comprising anchor nodes and mobile nodes distributed in two regions: a near zone (close to the anchors) and a far zone (further away from the anchors). We focus on the scenario where a small number of mobile nodes in the far zone have corrupted distance measurements. The challenge is to identify these susceptible nodes and robustly estimate their positions. To tackle this, we proposed dedicated algorithms for both identification and robust localization, supported by theoretical analysis that delineates conditions for exact corrupted node identification and precise position estimation. Additionally, we develop an algorithm to generate optimal anchor node configurations, ensuring maximal robustness in target node estimation.

Our numerical experiments on synthetic data, along with comparisons to state-of-the-art methods, demonstrate that the proposed algorithm effectively identifies susceptible nodes and achieves robust localization with a modest number of anchors. Future work will explore two directions. First, we plan to study structured noise modeled as $\bmm = \bX\bq + \bm{H}\bm{u}$, where $\bm{H}\bm{u}$ represents the outliers and $\bm{H}$ is the noise matrix, potentially inferred from historical measurements using learning algorithms or existing noise models. We aim to leverage connections to structured dictionary learning for this problem \citep{tasissa2024discriminative}. Second, we intend to incorporate a limited number of inter-mobile node distance measurements, particularly between nearby nodes, to assess whether such additional information can further improve robustness while minimizing the number of anchors.

\section*{Acknowledgment}
Abiy Tasissa acknowledge partial support from the National Science Foundation through grant DMS-2208392. 
\appendix
\section{Proof of Theorem \ref{thm:coherence_min_theorem}
}
\begin{proof}
    
Let $\bG=\bU\bm{\Sigma}\bV^T$ denote the singular value decomposition (SVD) of $\bG$, where $\bU\in \real^{s\times s}$ is an orthonormal matrix, $\bm{\Sigma}\in \real^{s\times t}$ is a rectangular diagonal matrix which contains its singular values along its diagonal, and $\bV \in \real^{t\times t}$ is an orthonormal matrix. Let $\sigma_1,\sigma_2,...,\sigma_s$ denote the singular values of $\bG$. Substituting the SVD of $\bG$ into the objective function of \eqref{eq:coherence_minimize}, we obtain
\begin{align*}
& \frac{1}{2}\|\bV\bm{\Sigma}^T\bU^T\bU\bm{\Sigma}\bV^T-\bm{I}\|_F^2\\
=& \frac{1}{2}\|\bV(\bm{\Sigma}^T\bm{\Sigma}-\bm{I})\bV^T\|_F^2\\
=&\frac{1}{2}\|\bm{\Sigma}^T\bm{\Sigma}-\bm{I}\|_F^2\\
=& \sum_{i=1}^{m} (\sigma_i^2-1)+\sum_{i=m+1}^{n} 1.
\end{align*}
Above, the first equality follows from  $\bU^T\bU=\bm{I}$ and $\bm{V}\bm{V}^T=\bm{I}$, while the second equality uses the invariance of the Frobenius norm under orthogonal transformation. To achieve the optimal solution, we set $\sigma_i=1$ for $i=1,\dots,..,m$. 
Substituting this back into the SVD yields the optimal solution $\bG_* = \bU\bV^T$, where $\bU\in \real^{s\times s}$ and $\bV\in \real^{s\times t}$.
\end{proof}

\bibliographystyle{IEEEtranN}
\bibliography{IEEEabrv,Robust_Node_Localization}

\begin{thebibliography}{56}
\providecommand{\natexlab}[1]{#1}
\providecommand{\url}[1]{#1}
\csname url@samestyle\endcsname
\providecommand{\newblock}{\relax}
\providecommand{\bibinfo}[2]{#2}
\providecommand{\BIBentrySTDinterwordspacing}{\spaceskip=0pt\relax}
\providecommand{\BIBentryALTinterwordstretchfactor}{4}
\providecommand{\BIBentryALTinterwordspacing}{\spaceskip=\fontdimen2\font plus
\BIBentryALTinterwordstretchfactor\fontdimen3\font minus
  \fontdimen4\font\relax}
\providecommand{\BIBforeignlanguage}[2]{{%
\expandafter\ifx\csname l@#1\endcsname\relax
\typeout{** WARNING: IEEEtranN.bst: No hyphenation pattern has been}%
\typeout{** loaded for the language `#1'. Using the pattern for}%
\typeout{** the default language instead.}%
\else
\language=\csname l@#1\endcsname
\fi
#2}}
\providecommand{\BIBdecl}{\relax}
\BIBdecl

\bibitem[Dargie et~al.(2023)Dargie, Wen, Panes-Ruiz, Riemenschneider,
  Ibarlucea, and Cuniberti]{dargie2023monitoring}
W.~Dargie, J.~Wen, L.~A. Panes-Ruiz, L.~Riemenschneider, B.~Ibarlucea, and
  G.~Cuniberti, ``Monitoring toxic gases using nanotechnology and wireless
  sensor networks,'' \emph{IEEE Sensors Journal}, vol.~23, no.~11, pp.
  12\,274--12\,283, 2023.

\bibitem[Sanjeevi et~al.(2020)Sanjeevi, Prasanna, Siva~Kumar, Gunasekaran,
  Alagiri, and Vijay~Anand]{sanjeevi2020precision}
P.~Sanjeevi, S.~Prasanna, B.~Siva~Kumar, G.~Gunasekaran, I.~Alagiri, and
  R.~Vijay~Anand, ``Precision agriculture and farming using internet of things
  based on wireless sensor network,'' \emph{Transactions on Emerging
  Telecommunications Technologies}, vol.~31, no.~12, p. e3978, 2020.

\bibitem[Wang et~al.(2023)Wang, Xing, Liu, and Wu]{Wang10054103}
R.~Wang, Z.~Xing, E.~Liu, and J.~Wu, ``Joint localization and communication
  study for intelligent reflecting surface aided wireless communication
  system,'' \emph{IEEE Transactions on Communications}, vol.~71, no.~5, pp.
  3024--3042, 2023.

\bibitem[Weiss et~al.(2022)Weiss, Arikan, Vishnu, Deane, Singer, and
  Wornell]{weiss2022semi}
A.~Weiss, T.~Arikan, H.~Vishnu, G.~B. Deane, A.~C. Singer, and G.~W. Wornell,
  ``A semi-blind method for localization of underwater acoustic sources,''
  \emph{IEEE Transactions on Signal Processing}, vol.~70, pp. 3090--3106, 2022.

\bibitem[Hoydis et~al.(2023)Hoydis, Aoudia, Cammerer, Nimier-David, Binder,
  Marcus, and Keller]{hoydis2023sionna}
J.~Hoydis, F.~A. Aoudia, S.~Cammerer, M.~Nimier-David, N.~Binder, G.~Marcus,
  and A.~Keller, ``Sionna rt: Differentiable ray tracing for radio propagation
  modeling,'' in \emph{2023 IEEE Globecom Workshops (GC Wkshps)}.\hskip 1em
  plus 0.5em minus 0.4em\relax IEEE, 2023, pp. 317--321.

\bibitem[Wang et~al.(2018)Wang, Zhou, Li, Sun, Wu, Jin, Quek, and
  Xu]{wang2018wireless}
J.~Wang, H.~Zhou, Y.~Li, Q.~Sun, Y.~Wu, S.~Jin, T.~Q. Quek, and C.~Xu,
  ``Wireless channel models for maritime communications,'' \emph{IEEE access},
  vol.~6, pp. 68\,070--68\,088, 2018.

\bibitem[Lei and Rice(2009)]{lei2009multipath}
Q.~Lei and M.~Rice, ``Multipath channel model for over-water aeronautical
  telemetry,'' \emph{IEEE Transactions on Aerospace and Electronic Systems},
  vol.~45, no.~2, pp. 735--742, 2009.

\bibitem[Shamaei and Kassas(2021)]{shamaei2021receiver}
K.~Shamaei and Z.~M. Kassas, ``Receiver design and time of arrival estimation
  for opportunistic localization with 5g signals,'' \emph{IEEE Transactions on
  Wireless Communications}, vol.~20, no.~7, pp. 4716--4731, 2021.

\bibitem[Zhao et~al.(2021)Zhao, Panerati, and Schoellig]{zhao2021learning}
W.~Zhao, J.~Panerati, and A.~P. Schoellig, ``Learning-based bias correction for
  time difference of arrival ultra-wideband localization of
  resource-constrained mobile robots,'' \emph{IEEE Robotics and Automation
  Letters}, vol.~6, no.~2, pp. 3639--3646, 2021.

\bibitem[Candes and Randall(2008)]{candes2008highly}
E.~J. Candes and P.~A. Randall, ``Highly robust error correction byconvex
  programming,'' \emph{IEEE Transactions on Information Theory}, vol.~54,
  no.~7, pp. 2829--2840, 2008.

\bibitem[Candes et~al.(2005{\natexlab{a}})Candes, Rudelson, Tao, and
  Vershynin]{candes2005error}
E.~Candes, M.~Rudelson, T.~Tao, and R.~Vershynin, ``Error correction via linear
  programming,'' in \emph{46th Annual IEEE Symposium on Foundations of Computer
  Science (FOCS'05)}.\hskip 1em plus 0.5em minus 0.4em\relax IEEE, 2005, pp.
  668--681.

\bibitem[Eldar et~al.(2010)Eldar, Kuppinger, and Bolcskei]{eldar2010block}
Y.~C. Eldar, P.~Kuppinger, and H.~Bolcskei, ``Block-sparse signals: Uncertainty
  relations and efficient recovery,'' \emph{IEEE Transactions on Signal
  Processing}, vol.~58, no.~6, pp. 3042--3054, 2010.

\bibitem[Baraniuk et~al.(2010)Baraniuk, Cevher, Duarte, and
  Hegde]{baraniuk2010model}
R.~G. Baraniuk, V.~Cevher, M.~F. Duarte, and C.~Hegde, ``Model-based
  compressive sensing,'' \emph{IEEE Transactions on information theory},
  vol.~56, no.~4, pp. 1982--2001, 2010.

\bibitem[Cand{\`e}s et~al.(2011)Cand{\`e}s, Li, Ma, and
  Wright]{candes2011robust}
E.~J. Cand{\`e}s, X.~Li, Y.~Ma, and J.~Wright, ``Robust principal component
  analysis?'' \emph{Journal of the ACM (JACM)}, vol.~58, no.~3, pp. 1--37,
  2011.

\bibitem[Dokmanic et~al.(2015)Dokmanic, Parhizkar, Ranieri, and
  Vetterli]{dokmanic2015euclidean}
I.~Dokmanic, R.~Parhizkar, J.~Ranieri, and M.~Vetterli, ``Euclidean distance
  matrices: essential theory, algorithms, and applications,'' \emph{IEEE Signal
  Processing Magazine}, vol.~32, no.~6, pp. 12--30, 2015.

\bibitem[Tang and Nehorai(2011)]{tang2011robust}
G.~Tang and A.~Nehorai, ``Robust principal component analysis based on low-rank
  and block-sparse matrix decomposition,'' in \emph{2011 45th Annual Conference
  on Information Sciences and Systems}.\hskip 1em plus 0.5em minus 0.4em\relax
  IEEE, 2011, pp. 1--5.

\bibitem[Liu et~al.(2015)Liu, Zhao, Yao, and Qi]{liu2015background}
X.~Liu, G.~Zhao, J.~Yao, and C.~Qi, ``Background subtraction based on low-rank
  and structured sparse decomposition,'' \emph{IEEE Transactions on Image
  Processing}, vol.~24, no.~8, pp. 2502--2514, 2015.

\bibitem[Zaeemzadeh et~al.(2017)Zaeemzadeh, Joneidi, Shahrasbi, and
  Rahnavard]{zaeemzadeh2017robust}
A.~Zaeemzadeh, M.~Joneidi, B.~Shahrasbi, and N.~Rahnavard, ``Robust target
  localization based on squared range iterative reweighted least squares,'' in
  \emph{2017 IEEE 14th International Conference on Mobile Ad Hoc and Sensor
  Systems (MASS)}.\hskip 1em plus 0.5em minus 0.4em\relax IEEE, 2017, pp.
  380--388.

\bibitem[Huber(2011)]{Huber2011}
\BIBentryALTinterwordspacing
P.~J. Huber, \emph{Robust Statistics}.\hskip 1em plus 0.5em minus 0.4em\relax
  Berlin, Heidelberg: Springer Berlin Heidelberg, 2011, pp. 1248--1251.
  [Online]. Available: \url{https://doi.org/10.1007/978-3-642-04898-2_594}
\BIBentrySTDinterwordspacing

\bibitem[Daubechies et~al.(2010)Daubechies, DeVore, Fornasier, and
  G{\"u}nt{\"u}rk]{daubechies2010iteratively}
I.~Daubechies, R.~DeVore, M.~Fornasier, and C.~S. G{\"u}nt{\"u}rk,
  ``Iteratively reweighted least squares minimization for sparse recovery,''
  \emph{Communications on Pure and Applied Mathematics: A Journal Issued by the
  Courant Institute of Mathematical Sciences}, vol.~63, no.~1, pp. 1--38, 2010.

\bibitem[Li et~al.(2017)Li, Ding, and Li]{li2017outlier}
X.~Li, S.~Ding, and Y.~Li, ``Outlier suppression via non-convex robust pca for
  efficient localization in wireless sensor networks,'' \emph{IEEE Sensors
  Journal}, vol.~17, no.~21, pp. 7053--7063, 2017.

\bibitem[Zhang et~al.(2023)Zhang, Tan, Ding, Li, and Li]{zhang2023device}
K.~Zhang, B.~Tan, S.~Ding, Y.~Li, and G.~Li, ``Device-free indoor localization
  based on sparse coding with nonconvex regularization and adaptive relaxation
  localization criteria,'' \emph{International Journal of Machine Learning and
  Cybernetics}, vol.~14, no.~2, pp. 429--443, 2023.

\bibitem[Youssef et~al.(2007)Youssef, Mah, and Agrawala]{youssef2007challenges}
M.~Youssef, M.~Mah, and A.~Agrawala, ``Challenges: device-free passive
  localization for wireless environments,'' in \emph{Proceedings of the 13th
  annual ACM international conference on Mobile computing and networking},
  2007, pp. 222--229.

\bibitem[Feng et~al.(2011)Feng, Au, Valaee, and Tan]{feng2011received}
C.~Feng, W.~S.~A. Au, S.~Valaee, and Z.~Tan, ``Received-signal-strength-based
  indoor positioning using compressive sensing,'' \emph{IEEE Transactions on
  mobile computing}, vol.~11, no.~12, pp. 1983--1993, 2011.

\bibitem[Moore et~al.(2004)Moore, Leonard, Rus, and Teller]{moore2004robust}
D.~Moore, J.~Leonard, D.~Rus, and S.~Teller, ``Robust distributed network
  localization with noisy range measurements,'' in \emph{Proceedings of the 2nd
  international conference on Embedded networked sensor systems}, 2004, pp.
  50--61.

\bibitem[Liberti et~al.(2014)Liberti, Lavor, Maculan, and
  Mucherino]{liberti2014euclidean}
L.~Liberti, C.~Lavor, N.~Maculan, and A.~Mucherino, ``Euclidean distance
  geometry and applications,'' \emph{Siam Review}, vol.~56, no.~1, pp. 3--69,
  2014.

\bibitem[Tasissa and Lai(2018)]{tasissa2018exact}
A.~Tasissa and R.~Lai, ``Exact reconstruction of euclidean distance geometry
  problem using low-rank matrix completion,'' \emph{IEEE Transactions on
  Information Theory}, vol.~65, no.~5, pp. 3124--3144, 2018.

\bibitem[Biswas et~al.(2006)Biswas, Lian, Wang, and Ye]{biswas2006semidefinite}
P.~Biswas, T.-C. Lian, T.-C. Wang, and Y.~Ye, ``Semidefinite programming based
  algorithms for sensor network localization,'' \emph{ACM Transactions on
  Sensor Networks (TOSN)}, vol.~2, no.~2, pp. 188--220, 2006.

\bibitem[Dargie and Poellabauer(2010)]{dargie2010fundamentals}
W.~Dargie and C.~Poellabauer, \emph{Fundamentals of wireless sensor networks:
  theory and practice}.\hskip 1em plus 0.5em minus 0.4em\relax John Wiley \&
  Sons, 2010.

\bibitem[Wang(2015)]{wang2015linear}
Y.~Wang, ``Linear least squares localization in sensor networks,''
  \emph{Eurasip journal on wireless communications and networking}, vol. 2015,
  pp. 1--7, 2015.

\bibitem[Lichtenberg and Tasissa(2024)]{lichtenberg2024localization}
S.~Lichtenberg and A.~Tasissa, ``Localization from structured distance matrices
  via low-rank matrix recovery,'' \emph{IEEE Transactions on Information
  Theory}, 2024.

\bibitem[Natarajan(1995)]{natarajan1995sparse}
B.~K. Natarajan, ``Sparse approximate solutions to linear systems,'' \emph{SIAM
  journal on computing}, vol.~24, no.~2, pp. 227--234, 1995.

\bibitem[Tropp(2004)]{tropp2004greed}
J.~A. Tropp, ``Greed is good: Algorithmic results for sparse approximation,''
  \emph{IEEE Transactions on Information theory}, vol.~50, no.~10, pp.
  2231--2242, 2004.

\bibitem[Chen et~al.(2001)Chen, Donoho, and Saunders]{chen2001atomic}
S.~S. Chen, D.~L. Donoho, and M.~A. Saunders, ``Atomic decomposition by basis
  pursuit,'' \emph{SIAM review}, vol.~43, no.~1, pp. 129--159, 2001.

\bibitem[Davis et~al.(1997)Davis, Mallat, and Avellaneda]{davis1997adaptive}
G.~Davis, S.~Mallat, and M.~Avellaneda, ``Adaptive greedy approximations,''
  \emph{Constructive approximation}, vol.~13, no.~1, pp. 57--98, 1997.

\bibitem[Pati et~al.(1993)Pati, Rezaiifar, and
  Krishnaprasad]{pati1993orthogonal}
Y.~C. Pati, R.~Rezaiifar, and P.~S. Krishnaprasad, ``Orthogonal matching
  pursuit: Recursive function approximation with applications to wavelet
  decomposition,'' in \emph{Proceedings of 27th Asilomar conference on signals,
  systems and computers}.\hskip 1em plus 0.5em minus 0.4em\relax IEEE, 1993,
  pp. 40--44.

\bibitem[Tropp and Gilbert(2007)]{tropp2007signal}
J.~A. Tropp and A.~C. Gilbert, ``Signal recovery from random measurements via
  orthogonal matching pursuit,'' \emph{IEEE Transactions on information
  theory}, vol.~53, no.~12, pp. 4655--4666, 2007.

\bibitem[Candes and Tao(2005)]{candes2005decoding}
E.~J. Candes and T.~Tao, ``Decoding by linear programming,'' \emph{IEEE
  transactions on information theory}, vol.~51, no.~12, pp. 4203--4215, 2005.

\bibitem[Foucart and Rauhut(2013)]{foucart2013mathematical}
\BIBentryALTinterwordspacing
S.~Foucart and H.~Rauhut, \emph{A Mathematical Introduction to Compressive
  Sensing}, ser. Applied and Numerical Harmonic Analysis.\hskip 1em plus 0.5em
  minus 0.4em\relax Springer New York, 2013. [Online]. Available:
  \url{https://books.google.com/books?id=zb28BAAAQBAJ}
\BIBentrySTDinterwordspacing

\bibitem[Donoho and Elad(2003)]{donoho2003optimally}
D.~L. Donoho and M.~Elad, ``Optimally sparse representation in general
  (nonorthogonal) dictionaries via $\ell_1$ minimization,'' \emph{Proc. the
  National Academy of Sciences}, vol. 100, no.~5, pp. 2197--2202, 2003.

\bibitem[Gribonval and Nielsen(2003)]{gribonval2003sparse}
R.~Gribonval and M.~Nielsen, ``Sparse representations in unions of bases,''
  \emph{IEEE transactions on Information theory}, vol.~49, no.~12, pp.
  3320--3325, 2003.

\bibitem[Trefethen and Bau(2022)]{trefethen2022numerical}
L.~N. Trefethen and D.~Bau, \emph{Numerical linear algebra}.\hskip 1em plus
  0.5em minus 0.4em\relax SIAM, 2022.

\bibitem[Monteiro and Adler(1989)]{monteiro1989interior}
R.~D. Monteiro and I.~Adler, ``Interior path following primal-dual algorithms.
  part i: Linear programming,'' \emph{Mathematical programming}, vol.~44,
  no.~1, pp. 27--41, 1989.

\bibitem[Yang et~al.(2013)Yang, Zhou, Balasubramanian, Sastry, and
  Ma]{yang2013fast}
A.~Y. Yang, Z.~Zhou, A.~G. Balasubramanian, S.~S. Sastry, and Y.~Ma, ``Fast
  {$\ell_1$}-minimization algorithms for robust face recognition,'' \emph{IEEE
  Transactions on Image Processing}, vol.~22, no.~8, pp. 3234--3246, 2013.

\bibitem[Beck and Teboulle(2009)]{beck2009fast}
A.~Beck and M.~Teboulle, ``A fast iterative shrinkage-thresholding algorithm
  for linear inverse problems,'' \emph{SIAM journal on imaging sciences},
  vol.~2, no.~1, pp. 183--202, 2009.

\bibitem[Candes et~al.(2005{\natexlab{b}})Candes, Romberg,
  et~al.]{candes2005l1}
E.~Candes, J.~Romberg \emph{et~al.}, ``l1-magic: Recovery of sparse signals via
  convex programming,'' \emph{URL: www. acm. caltech.
  edu/l1magic/downloads/l1magic. pdf}, vol.~4, no.~14, p.~16, 2005.

\bibitem[Welch(1967)]{welch1967lower}
L.~Welch, ``Lower bounds on the maximum cross correlation os signals,''
  \emph{IEEE Trans. Inf. Theory}, vol.~13, pp. 619--621, 1967.

\bibitem[Strohmer and Heath~Jr(2003)]{strohmer2003grassmannian}
T.~Strohmer and R.~W. Heath~Jr, ``Grassmannian frames with applications to
  coding and communication,'' \emph{Applied and computational harmonic
  analysis}, vol.~14, no.~3, pp. 257--275, 2003.

\bibitem[Jasper et~al.(2013)Jasper, Mixon, and Fickus]{jasper2013kirkman}
J.~Jasper, D.~G. Mixon, and M.~Fickus, ``Kirkman equiangular tight frames and
  codes,'' \emph{IEEE transactions on information theory}, vol.~60, no.~1, pp.
  170--181, 2013.

\bibitem[Fickus et~al.(2012)Fickus, Mixon, and Tremain]{fickus2012steiner}
M.~Fickus, D.~G. Mixon, and J.~C. Tremain, ``Steiner equiangular tight
  frames,'' \emph{Linear algebra and its applications}, vol. 436, no.~5, pp.
  1014--1027, 2012.

\bibitem[Grant and Boyd(2014)]{cvx}
M.~Grant and S.~Boyd, ``{CVX}: Matlab software for disciplined convex
  programming, version 2.1,'' \url{http://cvxr.com/cvx}, Mar. 2014.

\bibitem[Grant and Boyd(2008)]{gb08}
------, ``Graph implementations for nonsmooth convex programs,'' in
  \emph{Recent Advances in Learning and Control}, ser. Lecture Notes in Control
  and Information Sciences, V.~Blondel, S.~Boyd, and H.~Kimura, Eds.\hskip 1em
  plus 0.5em minus 0.4em\relax Springer-Verlag Limited, 2008, pp. 95--110,
  \url{http://stanford.edu/~boyd/graph_dcp.html}.

\bibitem[ApS(2012)]{mosek}
\BIBentryALTinterwordspacing
M.~ApS, \emph{Using MOSEK with CVX}, 2012. [Online]. Available:
  \url{https://cvxr.com/cvx/doc/mosek.html}
\BIBentrySTDinterwordspacing

\bibitem[Savvides et~al.(2003)Savvides, Garber, Adlakha, Moses, and
  Srivastava]{savvides2003error}
A.~Savvides, W.~Garber, S.~Adlakha, R.~Moses, and M.~B. Srivastava, ``On the
  error characteristics of multihop node localization in ad-hoc sensor
  networks,'' in \emph{Information Processing in Sensor Networks: Second
  International Workshop, IPSN 2003, Palo Alto, CA, USA, April 22--23, 2003
  Proceedings}.\hskip 1em plus 0.5em minus 0.4em\relax Springer, 2003, pp.
  317--332.

\bibitem[Zaeemzadeh(2020)]{zaeemzadeh2024robustlocalization}
\BIBentryALTinterwordspacing
A.~Zaeemzadeh, ``Robust\_localization,'' 2020, accessed: February 21, 2025.
  [Online]. Available: \url{https://github.com/zaeemzadeh/Robust_Localization}
\BIBentrySTDinterwordspacing

\bibitem[Tasissa et~al.(2024)Tasissa, Theodosis, Tolooshams, and
  Ba]{tasissa2024discriminative}
\BIBentryALTinterwordspacing
A.~Tasissa, M.~Theodosis, B.~Tolooshams, and D.~E. Ba, ``Discriminative
  reconstruction via simultaneous dense and sparse coding,'' \emph{Transactions
  on Machine Learning Research}, 2024. [Online]. Available:
  \url{https://openreview.net/forum?id=FkgM06HEbk}
\BIBentrySTDinterwordspacing

\end{thebibliography}

\end{document}